\documentclass[12pt]{article}
\usepackage{axodraw}
\begin{document}
%
\newcommand{\nl}{\nonumber\\}
\newcommand{\nn}{\nonumber}
\newcommand{\ds}{\displaystyle}
\newcommand{\mpar}[1]{{\marginpar{\hbadness10000%
                      \sloppy\hfuzz10pt\boldmath\bf#1}}%
                      \typeout{marginpar: #1}\ignorespaces}
\def\mnew{\mpar{\hfil NEW \hfil}\ignorespaces}
\newcommand{\lpar}{\left(}                            
\newcommand{\rpar}{\right)}
\newcommand{\lrbr}{\left[}
\newcommand{\rrbr}{\right]}
\newcommand{\lcbr}{\left\{}
\newcommand{\rcbr}{\right\}}
\newcommand{\rbrak}[1]{\lrbr#1\rrbr}
\newcommand{\bq}{\begin{equation}}                    
\newcommand{\eq}{\end{equation}}
\newcommand{\bqa}{\begin{eqnarray}}
\newcommand{\eqa}{\end{eqnarray}}
\newcommand{\ba}[1]{\begin{array}{#1}}
\newcommand{\ea}{\end{array}}
\newcommand{\ben}{\begin{enumerate}}
\newcommand{\een}{\end{enumerate}}
\newcommand{\bei}{\begin{itemize}}
\newcommand{\eei}{\end{itemize}}
\newcommand{\eqn}[1]{Eq.~(\ref{#1})}
\newcommand{\eqns}[2]{Eqs.~(\ref{#1})--(\ref{#2})}
\newcommand{\eqnss}[1]{Eqs.~(\ref{#1})}
\newcommand{\eqnsc}[2]{Eqs.~(\ref{#1}) and (\ref{#2})}
\newcommand{\eqnst}[3]{Eqs.~(\ref{#1}), (\ref{#2}) and (\ref{#3})}
\newcommand{\eqnsf}[4]{Eqs.~(\ref{#1}), (\ref{#2}), (\ref{#3}) and (\ref{#4})}
\newcommand{\eqnsv}[5]{Eqs.~(\ref{#1}), (\ref{#2}), (\ref{#3}), (\ref{#4}) and (\ref{#5})}
\newcommand{\tbn}[1]{Tab.~\ref{#1}}
\newcommand{\tabn}[1]{Tab.~\ref{#1}}
\newcommand{\tbns}[2]{Tabs.~\ref{#1}--\ref{#2}}
\newcommand{\tabns}[2]{Tabs.~\ref{#1}--\ref{#2}}
\newcommand{\tbnsc}[2]{Tabs.~\ref{#1} and \ref{#2}}
\newcommand{\fig}[1]{Fig.~\ref{#1}}
\newcommand{\figs}[2]{Figs.~\ref{#1}--\ref{#2}}
\newcommand{\sect}[1]{Section~\ref{#1}}
\newcommand{\sects}[2]{Section~\ref{#1} and \ref{#2}}
\newcommand{\subsect}[1]{Subsection~\ref{#1}}
\newcommand{\appendx}[1]{Appendix~\ref{#1}}
\newcommand{\hsp}{\hspace{.5mm}}
\def\negs{\hspace{-0.26in}}
\def\negsh{\hspace{-0.13in}}
%
%
\newcommand{\TeV}{\mathrm{TeV}}
\newcommand{\GeV}{\mathrm{GeV}}
\newcommand{\MeV}{\mathrm{MeV}}
\newcommand{\nb}{\mathrm{nb}}
\newcommand{\pb}{\mathrm{pb}}
\newcommand{\fb}{\mathrm{fb}}
\def\Re{\mathop{\operator@font Re}\nolimits}
\def\Im{\mathop{\operator@font Im}\nolimits}
\newcommand{\ord}[1]{{\cal O}\lpar#1\rpar}
\newcommand{\group}{SU(2)\otimes U(1)}
\newcommand{\ib}{i}
\newcommand{\asums}[1]{\sum_{#1}}
\newcommand{\asumt}[2]{\sum_{#1}^{#2}}
\newcommand{\asum}[3]{\sum_{#1=#2}^{#3}}
%
%
\newcommand{\fer}{{\rm{fer}}}
\newcommand{\bos}{{\rm{bos}}}
\newcommand{\lep}{{l}}
\newcommand{\had}{{h}}
\newcommand{\gen}{{g}}
\newcommand{\dbl}{{d}}
\newcommand{\philone}{\phi}
\newcommand{\philoneb}{\phi_{0}}
\newcommand{\phiind}[1]{\phi_{#1}}
\newcommand{\gBi}[2]{B_{#1}^{#2}}
\newcommand{\gBn}[1]{B^{#1}}
%
%
\newcommand{\ph}{\gamma}
\newcommand{\ab}{A}
\newcommand{\abp}{A'}
\newcommand{\abr}{A^r}
\newcommand{\abb}{A^{0}}
\newcommand{\abi}[1]{A_{#1}}
\newcommand{\abpi}[1]{A'_{#1}}
\newcommand{\abri}[1]{A^r_{#1}}
\newcommand{\abbi}[1]{A^{0}_{#1}}
\newcommand{\wb}{W}
\newcommand{\wbi}[1]{W_{#1}}
\newcommand{\wbp}{W^{+}}
\newcommand{\wbm}{W^{-}}
\newcommand{\wbpm}{W^{\pm}}
\newcommand{\wbpi}[1]{W^{+}_{#1}}
\newcommand{\wbmi}[1]{W^{-}_{#1}}
\newcommand{\wbpmi}[1]{W^{\pm}_{#1}}
\newcommand{\wbmpi}[1]{W^{\mp}_{#1}}
\newcommand{\wbli}[1]{W^{[+}_{#1}}
\newcommand{\wbri}[1]{W^{-]}_{#1}}
\newcommand{\zb}{Z}
\newcommand{\zbi}[1]{Z_{#1}}
\newcommand{\vb}{V}
\newcommand{\vbi}[1]{V_{#1}}
\newcommand{\vbiv}[1]{V^{*}_{#1}}
\newcommand{\Pb}{P}
\newcommand{\Sb}{S}
\newcommand{\Bb}{B}
%
%
\newcommand{\vph}{\varphi}
\newcommand{\hk}{K}
\newcommand{\hKi}[1]{K_{#1}}
\newcommand{\hkg}{\phi}
\newcommand{\hkn}{\phi^{0}}
\newcommand{\hkp}{\phi^{+}}
\newcommand{\hkm}{\phi^{-}}
\newcommand{\hkpm}{\phi^{\pm}}
\newcommand{\hkmp}{\phi^{\mp}}
\newcommand{\hki}[1]{\phi^{#1}}
\newcommand{\hb}{H}
\newcommand{\hbi}[1]{H_{#1}}
\newcommand{\hkl}{\phi^{[+\cgfi\cgfi}}
\newcommand{\hkr}{\phi^{-]}}
%
%
\newcommand{\fpx}{X}
\newcommand{\fpy}{Y}
\newcommand{\fpxp}{X^+}
\newcommand{\fpxm}{X^-}
\newcommand{\fpxpm}{X^{\pm}}
\newcommand{\fpxi}[1]{X^{#1}}
\newcommand{\fpyZ}{Y^{\ssZ}}
\newcommand{\fpyA}{Y^{\ssA}}
\newcommand{\fpyG}{Y^{\ssG}}
\newcommand{\fpyZA}{Y_{\ssZ,\ssA}}
\newcommand{\fpbxi}[1]{{\overline{X}}^{#1}}
\newcommand{\fpbyZ}{{\overline{Y}}^{\ssZ}}
\newcommand{\fpbyA}{{\overline{Y}}^{\ssA}}
\newcommand{\fpbyZA}{{\overline{Y}}^{\ssZ,\ssA}}
%
%
\newcommand{\Flone}{F}
\newcommand{\fpsi}{\psi}
\newcommand{\fpsif}{\psi_f}
\newcommand{\fpsifp}{\psi'_f}
\newcommand{\fpsii}[1]{\psi^{#1}}
\newcommand{\fbpsif}{{\overline{\psi}_f}}
\newcommand{\fbpsifp}{{\overline{\psi}'_f}}
\newcommand{\fpsib}{\psi_{\sszero}}
\newcommand{\fpsir}{\psi^r}
\newcommand{\fpsiL}{\psi_{_L}}
\newcommand{\fpsiR}{\psi_{_R}}
\newcommand{\fpsiLi}[1]{\psi_{_L}^{#1}}
\newcommand{\fpsiRi}[1]{\psi_{_R}^{#1}}
\newcommand{\fpsiLbi}[1]{\psi_{_{0L}}^{#1}}
\newcommand{\fpsiRbi}[1]{\psi_{_{0R}}^{#1}}
\newcommand{\fpsiLR}{\psi_{_{L,R}}}
\newcommand{\fbpsi}{{\overline{\psi}}}
\newcommand{\fbpsii}[1]{{\overline{\psi}}^{#1}}
\newcommand{\fbpsir}{{\overline{\psi}}^r}
\newcommand{\fbpsiL}{{\overline{\psi}}_{_L}}
\newcommand{\fbpsiR}{{\overline{\psi}}_{_R}}
\newcommand{\fbpsiLi}[1]{\overline{\psi_{_L}}^{#1}}
\newcommand{\fbpsiRi}[1]{\overline{\psi_{_R}}^{#1}}
\newcommand{\FQED}[2]{F_{#1#2}}
\newcommand{\FQEDp}[2]{F'_{#1#2}}
\newcommand{\fe}{e}
\newcommand{\ff}{f}
\newcommand{\fep}{e^{+}}
\newcommand{\fem}{e^{-}}
\newcommand{\fepm}{e^{\pm}}
\newcommand{\fp}{f^{+}}
\newcommand{\fm}{f^{-}}
\newcommand{\fhp}{h^{+}}
\newcommand{\fhm}{h^{-}}
\newcommand{\fh}{h}
\newcommand{\flm}{\mu}
\newcommand{\flmp}{\mu^{+}}
\newcommand{\flmm}{\mu^{-}}
\newcommand{\fll}{l}
\newcommand{\fllp}{l^{+}}
\newcommand{\fllm}{l^{-}}
\newcommand{\flt}{\tau}
\newcommand{\fltp}{\tau^{+}}
\newcommand{\fltm}{\tau^{-}}
\newcommand{\fq}{q}
\newcommand{\fqi}[1]{\fq_{#1}}
\newcommand{\bfqi}[1]{\barq_{#1}}
\newcommand{\ffQ}{Q}
\newcommand{\fu}{u}
\newcommand{\fd}{d}
\newcommand{\fc}{c}
\newcommand{\fs}{s}
\newcommand{\fqp}{q'}
\newcommand{\fup}{u'}
\newcommand{\fdp}{d'}
\newcommand{\fcp}{c'}
\newcommand{\fsp}{s'}
\newcommand{\fdpp}{d''}
\newcommand{\ffi}[1]{f_{#1}}
\newcommand{\bffi}[1]{{\overline{f}}_{#1}}
\newcommand{\ffpi}[1]{f'_{#1}}
\newcommand{\bffpi}[1]{{\overline{f}}'_{#1}}
\newcommand{\ft}{t}
\newcommand{\ffb}{b}
\newcommand{\ffp}{f'}
\newcommand{\fft}{{\tilde{f}}}
\newcommand{\fl}{l}
\newcommand{\fli}[1]{\fl_{#1}}
\newcommand{\fnu}{\nu}
\newcommand{\fU}{U}
\newcommand{\fD}{D}
\newcommand{\fUc}{\overline{U}}
\newcommand{\fDc}{\overline{D}}
\newcommand{\fnul}{\nu_l}
\newcommand{\fnue}{\nu_e}
\newcommand{\fnum}{\nu_{\mu}}
\newcommand{\fnut}{\nu_{\tau}}
\newcommand{\fbe}{{\overline{e}}}
\newcommand{\fblm}{{\overline{\mu}}}
\newcommand{\fblt}{{\overline{\tau}}}
\newcommand{\fbu}{{\overline{u}}}
\newcommand{\fbd}{{\overline{d}}}
\newcommand{\fbf}{{\overline{f}}}
\newcommand{\fbfp}{{\overline{f}}'}
\newcommand{\fbl}{{\overline{l}}}
\newcommand{\fbnu}{{\overline{\nu}}}
\newcommand{\fbnul}{{\overline{\nu}}_{\fl}}
\newcommand{\fbnue}{{\overline{\nu}}_{\fe}}
\newcommand{\fbnum}{{\overline{\nu}}_{\flm}}
\newcommand{\fbnut}{{\overline{\nu}}_{\flt}}
\newcommand{\fuL}{u_{_L}}
\newcommand{\fdL}{d_{_L}}
\newcommand{\ffL}{f_{_L}}
\newcommand{\fbuL}{{\overline{u}}_{_L}}
\newcommand{\fbdL}{{\overline{d}}_{_L}}
\newcommand{\fbfL}{{\overline{f}}_{_L}}
\newcommand{\fuR}{u_{_R}}
\newcommand{\fdR}{d_{_R}}
\newcommand{\ffR}{f_{_R}}
\newcommand{\fbuR}{{\overline{u}}_{_R}}
\newcommand{\fbdR}{{\overline{d}}_{_R}}
\newcommand{\fbfR}{{\overline{f}}_{_R}}
%
%
\newcommand{\barf}{\overline f}
\newcommand{\barl}{\overline l}
\newcommand{\barq}{\overline q}
\newcommand{\barqp}{\overline{q}'}
\newcommand{\barb}{\overline b}
\newcommand{\bart}{\overline t}
\newcommand{\barc}{\overline c}
\newcommand{\baru}{\overline u}
\newcommand{\bard}{\overline d}
\newcommand{\bars}{\overline s}
\newcommand{\barv}{\overline v}
\newcommand{\barnu}{\overline{\nu}}
\newcommand{\barne}{\overline{\nu}_{\fe}}
\newcommand{\barnm}{\overline{\nu}_{\flm}}
\newcommand{\barnt}{\overline{\nu}_{\flt}}
%
%
\newcommand{\glu}{g}
%
%
\newcommand{\phm}{\lambda}
\newcommand{\phms}{\lambda^2}
\newcommand{\mV}{M_{_V}}
\newcommand{\mw}{M_{_W}}
\newcommand{\mX}{M_{_X}}
\newcommand{\mY}{M_{_Y}}
\newcommand{\LM}{M}
\newcommand{\mz}{M_{_Z}}
\newcommand{\bzm}{M_{_0}}
\newcommand{\mh}{M_{_H}}
\newcommand{\bhm}{M_{_{0H}}}
\newcommand{\mf}{m_f}
\newcommand{\mfp}{m_{f'}}
\newcommand{\mfh}{m_{h}}
\newcommand{\mt}{m_t}
\newcommand{\me}{m_e}
\newcommand{\mm}{m_{\mu}}
\newcommand{\mtau}{m_{\tau}}
\newcommand{\muq}{m_u}
\newcommand{\md}{m_d}
\newcommand{\muqp}{m'_u}
\newcommand{\mdqp}{m'_d}
\newcommand{\mc}{m_c}
\newcommand{\ms}{m_s}
\newcommand{\mb}{m_b}
\newcommand{\mup}{M_u}                              
\newcommand{\mdp}{M_d}
\newcommand{\mcp}{M_c}
\newcommand{\msp}{M_s}
\newcommand{\mbp}{M_b}
%
%
\newcommand{\mls}{m^2_l}
\newcommand{\mVs}{M^2_{_V}}
\newcommand{\mVq}{M^4_{_V}}
\newcommand{\mws}{M^2_{_W}}
\newcommand{\mwc}{M^3_{_W}}
\newcommand{\LMs}{M^2}
\newcommand{\LMc}{M^3}
\newcommand{\mzs}{M^2_{_Z}}
\newcommand{\mzc}{M^3_{_Z}}
\newcommand{\bzms}{M^2_{_0}}
\newcommand{\bzmc}{M^3_{_0}}
\newcommand{\bhms}{M^2_{_{0H}}}
\newcommand{\mhs}{M^2_{_H}}
\newcommand{\mfs}{m^2_f}
\newcommand{\mfc}{m^3_f}
\newcommand{\mfps}{m^2_{f'}}
\newcommand{\mfhs}{m^2_{h}}
\newcommand{\mfpc}{m^3_{f'}}
\newcommand{\mts}{m^2_t}
\newcommand{\mes}{m^2_e}
\newcommand{\mms}{m^2_{\mu}}
\newcommand{\mmc}{m^3_{\mu}}
\newcommand{\mmfour}{m^4_{\mu}}
\newcommand{\mmf}{m^5_{\mu}}
\newcommand{\mmfive}{m^5_{\mu}}
\newcommand{\mmsix}{m^6_{\mu}}
\newcommand{\mtsix}{m^6_{\ft}}
\newcommand{\mminv}{\frac{1}{m_{\mu}}}
\newcommand{\mtaus}{m^2_{\tau}}
\newcommand{\mus}{m^2_u}
\newcommand{\mds}{m^2_d}
\newcommand{\muqps}{m'^2_u}
\newcommand{\mdqps}{m'^2_d}
\newcommand{\mcs}{m^2_c}
\newcommand{\mss}{m^2_s}
\newcommand{\mbs}{m^2_b}
\newcommand{\mups}{M^2_u}
\newcommand{\mdps}{M^2_d}
\newcommand{\mcps}{M^2_c}
\newcommand{\msps}{M^2_s}
\newcommand{\mbps}{M^2_b}
%
%
\newcommand{\mer}{m_{er}}
\newcommand{\mlep}{m_l}
\newcommand{\mleps}{m^2_l}
\newcommand{\mone}{m_1}
\newcommand{\mtwo}{m_2}
\newcommand{\mtre}{m_3}
\newcommand{\mfor}{m_4}
\newcommand{\mlone}{m}
\newcommand{\mloneb}{\bar{m}}
\newcommand{\mind}[1]{m_{#1}}
\newcommand{\mones}{m^2_1}
\newcommand{\mtwos}{m^2_2}
\newcommand{\mtres}{m^2_3}
\newcommand{\mfors}{m^2_4}
\newcommand{\mlones}{m^2}
\newcommand{\minds}[1]{m^2_{#1}}
\newcommand{\mlonec}{m^3}
\newcommand{\moneq}{m^4_1}
\newcommand{\mtwoq}{m^4_2}
\newcommand{\mtreq}{m^4_3}
\newcommand{\mforq}{m^4_4}
\newcommand{\mloneq}{m^4}
\newcommand{\mindq}[1]{m^4_{#1}}
\newcommand{\mlonev}{m^5}
\newcommand{\mindv}[1]{m^5_{#1}}
\newcommand{\monex}{m^6_1}
\newcommand{\mtwox}{m^6_2}
\newcommand{\mtrex}{m^6_3}
\newcommand{\mforx}{m^6_4}
\newcommand{\mlonex}{m^6}
\newcommand{\mindx}[1]{m^6_{#1}}
\newcommand{\Mone}{M_1}
\newcommand{\Mtwo}{M_2}
\newcommand{\Mtre}{M_3}
\newcommand{\Mfor}{M_4}
\newcommand{\Mlone}{M}
\newcommand{\Mlonep}{M'}
\newcommand{\Miind}{M_i}
\newcommand{\Mind}[1]{M_{#1}}
\newcommand{\Minds}[1]{M^2_{#1}}
\newcommand{\Mindc}[1]{M^3_{#1}}
\newcommand{\Mindf}[1]{M^4_{#1}}
\newcommand{\Mones}{M^2_1}
\newcommand{\Mtwos}{M^2_2}
\newcommand{\Mtres}{M^2_3}
\newcommand{\Mfors}{M^2_4}
\newcommand{\Mlones}{M^2}
\newcommand{\Mloneps}{M'^2}
\newcommand{\Miinds}{M^2_i}
\newcommand{\Mlonec}{M^3}
\newcommand{\Monec}{M^3_1}
\newcommand{\Mtwoc}{M^3_2}
\newcommand{\Moneq}{M^4_1}
\newcommand{\Mtwoq}{M^4_2}
\newcommand{\Mtreq}{M^4_3}
\newcommand{\Mforq}{M^4_4}
\newcommand{\Mloneq}{M^4}
\newcommand{\Miindq}{M^4_i}
\newcommand{\Monex}{M^6_1}
\newcommand{\Mtwox}{M^6_2}
\newcommand{\Mtrex}{M^6_3}
\newcommand{\Mforx}{M^6_4}
\newcommand{\Mlonex}{M^6}
\newcommand{\Miindx}{M^6_i}
\newcommand{\meb}{m_{\sszero}}
\newcommand{\mebs}{m^2_{\sszero}}
%
%
\newcommand{\Mq }{M_q  }
\newcommand{\MqS}{M^2_q}
\newcommand{\Ms }{M_s  }
\newcommand{\MsS}{M^2_s}
\newcommand{\Mc }{M_c  }
\newcommand{\McS}{M^2_c}
\newcommand{\Mb }{M_b  }
\newcommand{\MbS}{M^2_b}
\newcommand{\Mt }{M_t  }
\newcommand{\MtS}{M^2_t}
%
%
\newcommand{\mq}{m_q}
\newcommand{\mqs}{m^2_q}
\newcommand{\mqS}{m^2_q}
\newcommand{\mqQ}{m^4_q}
\newcommand{\mqX}{m^6_q}
\newcommand{\mqp}{m'_q }
\newcommand{\mqpS}{m'^2_q}
\newcommand{\mqpQ}{m'^4_q}
%
%
\newcommand{\lL}{l}
\newcommand{\lLi}[1]{l_{#1}}
\newcommand{\ls}{l^2}
\newcommand{\LL}{L}
\newcommand{\LcalL}{\cal{L}}
\newcommand{\LS}{L^2}
\newcommand{\LC}{L^3}
\newcommand{\LQ}{L^4}
\newcommand{\lw}{l_w}
\newcommand{\Lw}{L_w}
\newcommand{\Lws}{L^2_w}
\newcommand{\Lz}{L_z}
\newcommand{\Lzs}{L^2_z}
\newcommand{\Li}[1]{L_{#1}}
\newcommand{\Lis}[1]{L^2_{#1}}
\newcommand{\Lic}[1]{L^3_{#1}}
%
%
\newcommand{\sman}{s}
\newcommand{\tman}{t}
\newcommand{\uman}{u}
\newcommand{\smani}[1]{s_{#1}}
\newcommand{\bsmani}[1]{{\bar{s}}_{#1}}
\newcommand{\smans}{s^2}
\newcommand{\tmans}{t^2}
\newcommand{\umans}{u^2}
\newcommand{\shat}{{\hat s}}
\newcommand{\that}{{\hat t}}
\newcommand{\uhat}{{\hat u}}
%
%
\newcommand{\smanp}{s'}
\newcommand{\smanpi}[1]{s'_{#1}}
\newcommand{\tmanp}{t'}
\newcommand{\umanp}{u'}
\newcommand{\kappi}[1]{\kappa_{#1}}
\newcommand{\zetai}[1]{\zeta_{#1}}
%
%
%
\newcommand{\Phaspi}[1]{\Gamma_{#1}}
\newcommand{\rbetai}[1]{\beta_{#1}}
\newcommand{\ralphai}[1]{\alpha_{#1}}
\newcommand{\rbetais}[1]{\beta^2_{#1}}
\newcommand{\Lambdi}[1]{\Lambda_{#1}}
\newcommand{\Nomini}[1]{N_{#1}}
\newcommand{\smlone}{\frac{-\sman-\ib\ep}{\mlones}}
%
%
\newcommand{\theti}[1]{\theta_{#1}}
\newcommand{\delti}[1]{\delta_{#1}}
\newcommand{\phigi}[1]{\phi_{#1}}
\newcommand{\acoli}[1]{\xi_{#1}}
\newcommand{\scats}{s}
\newcommand{\scatss}{s^2}
\newcommand{\scatsi}[1]{s_{#1}}
\newcommand{\scatsis}[1]{s^2_{#1}}
\newcommand{\scatst}[2]{s_{#1}^{#2}}
\newcommand{\scatc}{c}
\newcommand{\scatcs}{c^2}
\newcommand{\scatci}[1]{c_{#1}}
\newcommand{\scatcis}[1]{c^2_{#1}}
\newcommand{\scatct}[2]{c_{#1}^{#2}}
\newcommand{\angamt}[2]{\gamma_{#1}^{#2}}
\newcommand{\scatsin}{\sin\theta}
\newcommand{\scatsins}{\sin^2\theta}
\newcommand{\scatcos}{\cos\theta}
\newcommand{\scatcoss}{\cos^2\theta}
%
%
\newcommand{\Regia}{{\cal{R}}}
\newcommand{\Iconi}[2]{{\cal{I}}_{#1}\lpar{#2}\rpar}
\newcommand{\sIcon}[1]{{\cal{I}}_{#1}}
\newcommand{\betaf}{\beta_{\ff}}
\newcommand{\betafs}{\beta^2_{\ff}}
\newcommand{\betafc}{\beta^3_{\ff}}
\newcommand{\Kfact}[2]{{\cal{K}}_{#1}\lpar{#2}\rpar}
%
%
\newcommand{\Struf}[4]{{\cal D}^{#1}_{#2}\lpar{#3;#4}\rpar}
\newcommand{\sStruf}[2]{{\cal D}^{#1}_{#2}}
\newcommand{\Fluxf}[2]{H\lpar{#1;#2}\rpar}
\newcommand{\Fluxfi}[4]{H_{#1}^{#2}\lpar{#3;#4}\rpar}
\newcommand{\sFluxf}{H}
\newcommand{\Bflux}[2]{{\cal{B}}_{#1}\lpar{#2}\rpar}
\newcommand{\bflux}[2]{{\cal{B}}_{#1}\lpar{#2}\rpar}
\newcommand{\Fluxd}[2]{D_{#1}\lpar{#2}\rpar}
\newcommand{\fluxd}[2]{C_{#1}\lpar{#2}\rpar}
\newcommand{\Fluxh}[4]{{\cal{H}}_{#1}^{#2}\lpar{#3;#4}\rpar}
\newcommand{\Sluxh}[4]{{\cal{S}}_{#1}^{#2}\lpar{#3;#4}\rpar}
\newcommand{\Fluxhb}[4]{{\overline{{\cal{H}}}}_{#1}^{#2}\lpar{#3;#4}\rpar}
\newcommand{\sFluxhb}{{\overline{{\cal{H}}}}}
\newcommand{\Sluxhb}[4]{{\overline{{\cal{S}}}}_{#1}^{#2}\lpar{#3;#4}\rpar}
\newcommand{\sSluxhb}[2]{{\overline{{\cal{S}}}}_{#1}^{#2}}
\newcommand{\fluxh}[4]{h_{#1}^{#2}\lpar{#3;#4}\rpar}
\newcommand{\fluxhs}[3]{h_{#1}^{#2}\lpar{#3}\rpar}
\newcommand{\sfluxhs}[2]{h_{#1}^{#2}}
\newcommand{\fluxhb}[4]{{\overline{h}}_{#1}^{#2}\lpar{#3;#4}\rpar}
\newcommand{\Strufd}[2]{D\lpar{#1;#2}\rpar}
%
%
\newcommand{\rMQ}[1]{r^2_{#1}}
\newcommand{\rMQs}[1]{r^4_{#1}}
\newcommand{\hf}{h_{\ff}}
\newcommand{\rf}{w_{\ff}}
\newcommand{\zf}{z_{\ff}}
\newcommand{\rfs}{w^2_{\ff}}
\newcommand{\zfs}{z^2_{\ff}}
\newcommand{\rfc}{w^3_{\ff}}
\newcommand{\zfc}{z^3_{\ff}}
\newcommand{\df}{d_{\ff}}
\newcommand{\rfp}{w_{\ffp}}
\newcommand{\rfps}{w^2_{\ffp}}
\newcommand{\rfpc}{w^3_{\ffp}}
\newcommand{\Lrfp}{L_{w}}
\newcommand{\rt}{w_{\ft}}
\newcommand{\rts}{w^2_{\ft}}
\newcommand{\rb}{w_{\ffb}}
\newcommand{\rbs}{w^2_{\ffb}}
\newcommand{\dt}{d_{\ft}}
\newcommand{\dts}{d^2_{\ft}}
\newcommand{\rh}{r_{h}}
\newcommand{\Lnrt}{\ln{\rt}}
\newcommand{\Rw}{R_{_{\wb}}}
\newcommand{\Rws}{R^2_{_{\wb}}}
\newcommand{\Rz}{R_{_{\zb}}}
\newcommand{\Rzp}{R^{+}_{_{\zb}}}
\newcommand{\Rzm}{R^{-}_{_{\zb}}}
\newcommand{\Rzs}{R^2_{_{\zb}}}
\newcommand{\Rzc}{R^3_{_{\zb}}}
\newcommand{\Rv}{R_{_{\vb}}}
\newcommand{\rhw}{w_h}
\newcommand{\rhz}{z_h}
\newcommand{\rhws}{w^2_h}
\newcommand{\rhzs}{z^2_h}
%
%
\newcommand{\vqrato}{z}
\newcommand{\vqrats}{w}
\newcommand{\vqratq}{w^2}
\newcommand{\seyrat}{z}
\newcommand{\sexrat}{w}
\newcommand{\sexrats}{w^2}
\newcommand{\sehrat}{h}
\newcommand{\sewrat}{w}
\newcommand{\sezrat}{z}
\newcommand{\zetav}{\zeta}
\newcommand{\zetavi}[1]{\zeta_{#1}}
\newcommand{\bpo}{\beta^2}
\newcommand{\bpos}{\beta^4}
\newcommand{\bpt}{{\tilde\beta}^2}
\newcommand{\lap}{\kappa}
\newcommand{\hw}{w_h}
\newcommand{\hz}{z_h}
%
%
\newcommand{\ec}{e}
\newcommand{\ecs}{e^2}
\newcommand{\ect}{e^3}
\newcommand{\ecq}{e^4}
\newcommand{\ecb}{e_{\sszero}}
\newcommand{\ecbs}{e^2_{_0}}
\newcommand{\ecbq}{e^4_{_0}}
\newcommand{\eci}[1]{e_{#1}}
\newcommand{\ecis}[1]{e^2_{#1}}
\newcommand{\hate}{{\hat e}}
\newcommand{\gss}{g_{_S}}
\newcommand{\gsss}{g^2_{_S}}
\newcommand{\gssb}{g^2_{_{S_0}}}
\newcommand{\als}{\alpha_{_S}}
\newcommand{\as}{a_{_S}}
\newcommand{\ass}{a^2_{_S}}
\newcommand{\gf}{G_{\ssF}}
\newcommand{\gfs}{G^2_{\ssF}}
\newcommand{\gb}{g}
\newcommand{\gbi}[1]{g_{#1}}
\newcommand{\gbb}{g_{0}}
\newcommand{\gbs}{g^2}
\newcommand{\gbc}{g^3}
\newcommand{\gbf}{g^4}
\newcommand{\gpb}{g'}
\newcommand{\gpbs}{g'^2}
\newcommand{\vc}[1]{v_{#1}}
\newcommand{\ac}[1]{a_{#1}}
\newcommand{\vcc}[1]{v^*_{#1}}
\newcommand{\acc}[1]{a^*_{#1}}
\newcommand{\hatv}[1]{{\hat v}_{#1}}
\newcommand{\vcs}[1]{v^2_{#1}}
\newcommand{\acs}[1]{a^2_{#1}}
\newcommand{\gcv}[1]{g^{#1}_{\ssV}}
\newcommand{\gca}[1]{g^{#1}_{\ssA}}
\newcommand{\gcp}[1]{g^{+}_{#1}}
\newcommand{\gcm}[1]{g^{-}_{#1}}
\newcommand{\gcpm}[1]{g^{\pm}_{#1}}
\newcommand{\vci}[2]{v^{#2}_{#1}}
\newcommand{\aci}[2]{a^{#2}_{#1}}
\newcommand{\vceff}[1]{v^{#1}_{\rm{eff}}}
\newcommand{\hvc}[1]{\hat{v}_{#1}}
\newcommand{\hvcs}[1]{\hat{v}^2_{#1}}
\newcommand{\Vc}[1]{V_{#1}}
\newcommand{\Ac}[1]{A_{#1}}
\newcommand{\Vcs}[1]{V^2_{#1}}
\newcommand{\Acs}[1]{A^2_{#1}}
\newcommand{\vpa}[2]{\sigma_{#1}^{#2}}
\newcommand{\vma}[2]{\delta_{#1}^{#2}}
\newcommand{\vfw}{\sigma^{a}_{\ff}}
\newcommand{\vfpw}{\sigma^{a}_{\ffp}}
\newcommand{\vfwi}[1]{\sigma^{a}_{#1}}
\newcommand{\vfwsi}[1]{\lpar\sigma^{a}_{#1}\rpar^2}
\newcommand{\vvfw}{v^{a}_{\ff}}
\newcommand{\vvew}{v^{a}_{\fe}}
\newcommand{\vzm}{v^{-}_{\ssZ}}
\newcommand{\vzp}{v^{+}_{\ssZ}}
\newcommand{\vzpm}{v^{\pm}_{\ssZ}}
\newcommand{\vzmp}{v^{\mp}_{\ssZ}}
\newcommand{\vam}{v^{-}_{\ssA}}
\newcommand{\vap}{v^{+}_{\ssA}}
\newcommand{\vapm}{v^{\pm}_{\ssA}}
\newcommand{\gv}{g_{_V}}
\newcommand{\ga}{g_{_A}}
\newcommand{\gve}{g^{\fe}_{_{V}}}
\newcommand{\gae}{g^{\fe}_{_{A}}}
\newcommand{\gvf}{g^{\ff}_{_{V}}}
\newcommand{\gaf}{g^{\ff}_{_{A}}}
\newcommand{\gva}{g_{_{V,A}}}
\newcommand{\gvae}{g^{\fe}_{_{V,A}}}
\newcommand{\gvaf}{g^{\ff}_{_{V,A}}}
\newcommand{\sGv}{{\cal{G}}_{_V}}
\newcommand{\cGa}{{\cal{G}}^{*}_{_A}}
\newcommand{\cGv}{{\cal{G}}^{*}_{_V}}
\newcommand{\sGa}{{\cal{G}}_{_A}}
\newcommand{\Gvf}{{\cal{G}}^{\ff}_{_{V}}}
\newcommand{\Gaf}{{\cal{G}}^{\ff}_{_{A}}}
\newcommand{\Gvaf}{{\cal{G}}^{\ff}_{_{V,A}}}
\newcommand{\Gve}{{\cal{G}}^{\fe}_{_{V}}}
\newcommand{\Gae}{{\cal{G}}^{\fe}_{_{A}}}
\newcommand{\Gvae}{{\cal{G}}^{\fe}_{_{V,A}}}
\newcommand{\gvl}{g^{\fl}_{_{V}}}
\newcommand{\gal}{g^{\fl}_{_{A}}}
\newcommand{\gval}{g^{\fl}_{_{V,A}}}
\newcommand{\gvb}{g^{\ffb}_{_{V}}}
\newcommand{\gab}{g^{\ffb}_{_{A}}}
\newcommand{\fvf}{F_{_V}^{\ff}}
\newcommand{\faf}{F_{_A}^{\ff}}
\newcommand{\fvl}{F_{_V}^{\fl}}
\newcommand{\fal}{F_{_A}^{\fl}}
\newcommand{\corat}{\kappa}
\newcommand{\corats}{\kappa^2}
%
%
\newcommand{\dr}{\Delta r}
\newcommand{\drl}{\Delta r_{_L}}
\newcommand{\drh}{\Delta{\hat r}}
\newcommand{\drhw}{\Delta{\hat r}_{_W}}
\newcommand{\rhou}{\rho_{_U}}
\newcommand{\rhoz}{\rho_{_\zb}}
\newcommand{\rZ}{\rho_{_\zb}}
\newcommand{\rhob}{\rho_{_0}}
\newcommand{\rZf}{\rho^{\ff}_{_\zb}}
\newcommand{\rhoe}{\rho_{\fe}}
\newcommand{\rhof}{\rho_{\ff}}
\newcommand{\rhoi}[1]{\rho_{#1}}
\newcommand{\kZf}{\kappa^{\ff}_{_\zb}}
\newcommand{\rWf}{\rho^{\ff}_{_\wb}}
\newcommand{\brWf}{{\bar{\rho}}^{\ff}_{_\wb}}
\newcommand{\rHf}{\rho^{\ff}_{_\hb}}
\newcommand{\brHf}{{\bar{\rho}}^{\ff}_{_\hb}}
\newcommand{\rhoR}{\rho^{\ssR}_{_{\zb}}}
\newcommand{\hatrh}{{\hat\rho}}
\newcommand{\ku}{\kappa_{\ssU}}
\newcommand{\rZdf}[1]{\rho^{#1}_{_\zb}}
\newcommand{\kZdf}[1]{\kappa^{#1}_{_\zb}}
\newcommand{\rdfL}[1]{\rho^{#1}_{_L}}
\newcommand{\kdfL}[1]{\kappa^{#1}_{_L}}
\newcommand{\rdfR}[1]{\rho^{#1}_{\rm{rem}}}
\newcommand{\kdfR}[1]{\kappa^{#1}_{\rm{rem}}}
\newcommand{\bark}{\overline\kappa}
%
%
\newcommand{\stw}{s_{\theta}}             
\newcommand{\ctw}{c_{\theta}}
\newcommand{\stws}{s_{\theta}^2}
\newcommand{\stwc}{s_{\theta}^3}
\newcommand{\stwf}{s_{\theta}^4}
\newcommand{\stwx}{s_{\theta}^6}
\newcommand{\ctws}{c_{\theta}^2}
\newcommand{\ctwc}{c_{\theta}^3}
\newcommand{\ctwf}{c_{\theta}^4}
\newcommand{\ctwx}{c_{\theta}^6}
\newcommand{\stwfiv}{s_{\theta}^5}
\newcommand{\ctwfiv}{c_{\theta}^5}
\newcommand{\stwsix}{s_{\theta}^6}
\newcommand{\ctwsix}{c_{\theta}^6}
%
%
\newcommand{\siw}{s_{_W}}
\newcommand{\cow}{c_{_W}}
\newcommand{\siws}{s^2_{_W}}
\newcommand{\cows}{c^2_{_W}}
\newcommand{\siwc}{s^3_{_W}}
\newcommand{\cowc}{c^3_{_W}}
\newcommand{\siwf}{s^4_{_W}}
\newcommand{\cowf}{c^4_{_W}}
\newcommand{\siwx}{s^6_{_W}}
\newcommand{\cowx}{c^6_{_W}}
\newcommand{\sons}{s_{_W}}
\newcommand{\sonss}{s^2_{_W}}
\newcommand{\cons}{c_{_W}}
\newcommand{\cooss}{c^2_{_W}}
%
%
\newcommand{\szs}{{\overline s}^2}
\newcommand{\szq}{{\overline s}^4}
\newcommand{\czs}{{\overline c}^2}
\newcommand{\sbs}{s_{_0}^2}
\newcommand{\cbs}{c_{_0}^2}
\newcommand{\dss}{\Delta s^2}
\newcommand{\snes}{s_{\nu e}^2}
\newcommand{\cnes}{c_{\nu e}^2}
\newcommand{\shs}{{\hat s}^2}
\newcommand{\chs}{{\hat c}^2}
\newcommand{\chl}{{\hat c}}
\newcommand{\seffs}{s^2_{\rm{eff}}}
\newcommand{\seffsf}[1]{\sin^2\theta^{#1}_{\rm{eff}}}
\newcommand{\sress}{s^2_{\rm res}}
\newcommand{\sR}{s_{_R}}
\newcommand{\sRs}{s^2_{_R}}
\newcommand{\ctwe}{c_{\theta}^6}
\newcommand{\sany}{s}
\newcommand{\cany}{c}
\newcommand{\sanys}{s^2}
\newcommand{\canys}{c^2}
%
%
\newcommand{\sip}{u}                             
\newcommand{\siap}{{\bar{v}}}                    
\newcommand{\sop}{{\bar{u}}}                     
\newcommand{\soap}{v}                            
\newcommand{\ip}[1]{u\lpar{#1}\rpar}             
\newcommand{\iap}[1]{{\bar{v}}\lpar{#1}\rpar}    
\newcommand{\op}[1]{{\bar{u}}\lpar{#1}\rpar}     
\newcommand{\oap}[1]{v\lpar{#1}\rpar}            
%
%
\newcommand{\ipp}[2]{u\lpar{#1,#2}\rpar}         
\newcommand{\ipap}[2]{{\bar v}\lpar{#1,#2}\rpar} 
\newcommand{\opp}[2]{{\bar u}\lpar{#1,#2}\rpar}  
\newcommand{\opap}[2]{v\lpar{#1,#2}\rpar}        
\newcommand{\upspi}[1]{u\lpar{#1}\rpar}
\newcommand{\vpspi}[1]{v\lpar{#1}\rpar}
\newcommand{\wpspi}[1]{w\lpar{#1}\rpar}
\newcommand{\ubpspi}[1]{{\bar{u}}\lpar{#1}\rpar}
\newcommand{\vbpspi}[1]{{\bar{v}}\lpar{#1}\rpar}
\newcommand{\wbpspi}[1]{{\bar{w}}\lpar{#1}\rpar}
\newcommand{\udpspi}[1]{u^{\dagger}\lpar{#1}\rpar}
\newcommand{\vdpspi}[1]{v^{\dagger}\lpar{#1}\rpar}
\newcommand{\wdpspi}[1]{w^{\dagger}\lpar{#1}\rpar}
\newcommand{\Ubilin}[1]{U\lpar{#1}\rpar}
\newcommand{\Vbilin}[1]{V\lpar{#1}\rpar}
\newcommand{\Xbilin}[1]{X\lpar{#1}\rpar}
\newcommand{\Ybilin}[1]{Y\lpar{#1}\rpar}
\newcommand{\up}[2]{u_{#1}\lpar #2\rpar}
\newcommand{\vp}[2]{v_{#1}\lpar #2\rpar}
\newcommand{\ubp}[2]{{\overline u}_{#1}\lpar #2\rpar}
\newcommand{\vbp}[2]{{\overline v}_{#1}\lpar #2\rpar}
\newcommand{\Pje}[1]{\frac{1}{2}\lpar 1 + #1\,\gfd\rpar}
\newcommand{\Pj}[1]{\Pi_{#1}}
\newcommand{\trace}{\mbox{Tr}}
%
%
\newcommand{\Poper}[2]{P_{#1}\lpar{#2}\rpar}
\newcommand{\Loper}[2]{\Lambda_{#1}\lpar{#2}\rpar}
\newcommand{\proj}[3]{P_{#1}\lpar{#2,#3}\rpar}
\newcommand{\sproj}[1]{P_{#1}}
\newcommand{\Nden}[3]{N_{#1}^{#2}\lpar{#3}\rpar}
\newcommand{\sNden}[1]{N_{#1}}
\newcommand{\nden}[2]{n_{#1}^{#2}}
%
%
\newcommand{\vwf}[2]{e_{#1}\lpar#2\rpar}             
\newcommand{\vwfb}[2]{{\overline e}_{#1}\lpar#2\rpar}
\newcommand{\pwf}[2]{\epsilon_{#1}\lpar#2\rpar}      
\newcommand{\sla}[1]{/\!\!\!#1}
\newcommand{\slac}[1]{/\!\!\!\!#1}
%
%
\newcommand{\iemom}{p_{_-}}                    
\newcommand{\ipmom}{p_{_+}}
\newcommand{\oemom}{q_{_-}}                    
\newcommand{\opmom}{q_{_+}}
%
%
\newcommand{\spro}[2]{{#1}\cdot{#2}}
%
%
\newcommand{\gfour}{\gamma_4}
\newcommand{\gfd}{\gamma_5}
\newcommand{\gap}{\lpar 1+\gamma_5\rpar}
\newcommand{\gam}{\lpar 1-\gamma_5\rpar}
\newcommand{\gdp}{\gamma_+}
\newcommand{\gdm}{\gamma_-}
\newcommand{\gdpm}{\gamma_{\pm}}
\newcommand{\gad}{\gamma}
\newcommand{\gapm}{\lpar 1\pm\gamma_5\rpar}
\newcommand{\gadi}[1]{\gamma_{#1}}
\newcommand{\gadu}[1]{\gamma_{#1}}
\newcommand{\gaduc}[1]{\gamma^{*}_{#1}}
\newcommand{\gaduh}[1]{\gamma^{+}_{#1}}
\newcommand{\gadut}[1]{\gamma^{\ssT}_{#1}}
\newcommand{\gapu}[1]{\gamma^{#1}}
\newcommand{\sigd}[2]{\sigma_{#1#2}}
\newcommand{\sumsp}{\overline{\sum_{\mbox{\tiny{spins}}}}}
%
%
\newcommand{\li}[2]{\mathrm{Li}_{#1}\lpar\displaystyle{#2}\rpar} 
\newcommand{\sli}[1]{\mathrm{Li}_{#1}} 
\newcommand{\etaf}[2]{\eta\lpar#1,#2\rpar}
\newcommand{\lkall}[3]{\lambda\lpar#1,#2,#3\rpar}       
\newcommand{\slkall}[3]{\lambda^{1/2}\lpar#1,#2,#3\rpar}
\newcommand{\segam}{\Gamma}                             
\newcommand{\egam}[1]{\Gamma\lpar#1\rpar}               
\newcommand{\egams}[1]{\Gamma^2\lpar#1\rpar}            
\newcommand{\ebe}[2]{B\lpar#1,#2\rpar}                  
\newcommand{\ddel}[1]{\delta\lpar#1\rpar}               
\newcommand{\ddeln}[1]{\delta^{(n)}\lpar#1\rpar}        
\newcommand{\drii}[2]{\delta_{#1#2}}                    
\newcommand{\driv}[4]{\delta_{#1#2#3#4}}                
\newcommand{\intmomi}[2]{\int\,d^{#1}#2}
\newcommand{\intmomii}[3]{\int\,d^{#1}#2\,\int\,d^{#1}#3}
\newcommand{\intfx}[1]{\int_{\scriptstyle 0}^{\scriptstyle 1}\,d#1}
\newcommand{\intfxy}[2]{\int_{\scriptstyle 0}^{\scriptstyle 1}\,d#1\,
                        \int_{\scriptstyle 0}^{\scriptstyle #1}\,d#2}
\newcommand{\intfxyz}[3]{\int_{\scriptstyle 0}^{\scriptstyle 1}\,d#1\,
                         \int_{\scriptstyle 0}^{\scriptstyle #1}\,d#2\,
                         \int_{\scriptstyle 0}^{\scriptstyle #2}\,d#3}
\newcommand{\Beta}[2]{{\rm{B}}\lpar #1,#2\rpar}
\newcommand{\sBeta}{\rm{B}}
\newcommand{\sign}[1]{{\rm{sign}}\lpar{#1}\rpar}
%
%
\newcommand{\gn}{\Gamma_{\nu}}
\newcommand{\gel}{\Gamma_{\fe}}
\newcommand{\gmu}{\Gamma_{\mu}}
\newcommand{\gff}{\Gamma_{\ff}}
\newcommand{\gt}{\Gamma_{\tau}}
\newcommand{\gl}{\Gamma_{\fl}}
\newcommand{\gq}{\Gamma_{\fq}}
\newcommand{\gu}{\Gamma_{\fu}}
\newcommand{\gd}{\Gamma_{\fd}}
\newcommand{\gc}{\Gamma_{\fc}}
\newcommand{\gs}{\Gamma_{\fs}}
\newcommand{\gbq}{\Gamma_{\ffb}}
\newcommand{\gz}{\Gamma_{_{\zb}}}
\newcommand{\gw}{\Gamma_{_{\wb}}}
\newcommand{\gh}{\Gamma_{\had}}
\newcommand{\ghb}{\Gamma_{_{\hb}}}
\newcommand{\gi}{\Gamma_{\rm{inv}}}
\newcommand{\gzs}{\Gamma^2_{_{\zb}}}
%
%
\newcommand{\tcie}{I^{(3)}_{\fe}}
\newcommand{\tcim}{I^{(3)}_{\flm}}
\newcommand{\tcif}{I^{(3)}_{\ff}}
\newcommand{\tciq}{I^{(3)}_{\fq}}
\newcommand{\tcib}{I^{(3)}_{\ffb}}
\newcommand{\tcih}{I^{(3)}_h}
\newcommand{\tcii}{I^{(3)}_i}
\newcommand{\tcift}{I^{(3)}_{\tilde f}}
\newcommand{\tcifp}{I^{(3)}_{f'}}
\newcommand{\wispt}[1]{I^{(3)}_{#1}}
\newcommand{\ql}{Q_l}
\newcommand{\qe}{Q_e}
\newcommand{\qu}{Q_u}
\newcommand{\qd}{Q_d}
\newcommand{\qb}{Q_b}
\newcommand{\qt}{Q_t}
\newcommand{\qup}{Q'_u}
\newcommand{\qdp}{Q'_d}
\newcommand{\qmu}{Q_{\mu}}
\newcommand{\qes}{Q^2_e}
\newcommand{\qec}{Q^3_e}
\newcommand{\qus}{Q^2_u}
\newcommand{\qds}{Q^2_d}
\newcommand{\qbs}{Q^2_b}
\newcommand{\qts}{Q^2_t}
\newcommand{\qbc}{Q^3_b}
\newcommand{\qf}{Q_f}
\newcommand{\qfs}{Q^2_f}
\newcommand{\qfc}{Q^3_f}
\newcommand{\qff}{Q^4_f}
\newcommand{\qep}{Q_{e'}}
\newcommand{\qfp}{Q_{f'}}
\newcommand{\qfps}{Q^2_{f'}}
\newcommand{\qfpc}{Q^3_{f'}}
\newcommand{\qq}{Q_q}
\newcommand{\qqs}{Q^2_q}
\newcommand{\qi}{Q_i}
\newcommand{\qis}{Q^2_i}
\newcommand{\qj}{Q_j}
\newcommand{\qjs}{Q^2_j}
\newcommand{\QW}{Q_{_\wb}}
\newcommand{\QWs}{Q^2_{_\wb}}
\newcommand{\Qd}{Q_d}
\newcommand{\Qds}{Q^2_d}
\newcommand{\Qu}{Q_u}
\newcommand{\Qus}{Q^2_u}
\newcommand{\vi}{v_i}
\newcommand{\vis}{v^2_i}
\newcommand{\ai}{a_i}
\newcommand{\ais}{a^2_i}
%
%
\newcommand{\piv}{\Pi_{_V}}
\newcommand{\pia}{\Pi_{_A}}
\newcommand{\piva}{\Pi_{_{V,A}}}
\newcommand{\pivi}[1]{\Pi^{({#1})}_{_V}}
\newcommand{\piai}[1]{\Pi^{({#1})}_{_A}}
\newcommand{\pivai}[1]{\Pi^{({#1})}_{_{V,A}}}
\newcommand{\pih}{{\hat\Pi}}
\newcommand{\sgh}{{\hat\Sigma}}
\newcommand{\Pgg}{\Pi_{\ph\ph}}
\newcommand{\Ptg}{\Pi_{_{3Q}}}
\newcommand{\Ptt}{\Pi_{_{33}}}
\newcommand{\Pzg}{\Pi_{_{\zb\ab}}}
\newcommand{\Pzga}[2]{\Pi^{#1}_{_{\zb\ab}}\lpar#2\rpar}
\newcommand{\Pf}{\Pi^{\ssF}}
\newcommand{\Sgg}{\Sigma_{_{\ab\ab}}}
\newcommand{\Szg}{\Sigma_{_{\zb\ab}}}
\newcommand{\SVV}{\Sigma_{_{\vb\vb}}}
\newcommand{\USvv}{{\hat\Sigma}_{_{\vb\vb}}}
\newcommand{\Sww}{\Sigma_{_{\wb\wb}}}
\newcommand{\Swwg}{\Sigma^{_G}_{_{\wb\wb}}}
\newcommand{\Szz}{\Sigma_{_{\zb\zb}}}
\newcommand{\Shh}{\Sigma_{_{\hb\hb}}}
\newcommand{\Spzz}{\Sigma'_{_{\zb\zb}}}
\newcommand{\Stg}{\Sigma_{_{3Q}}}
\newcommand{\Stt}{\Sigma_{_{33}}}
\newcommand{\bSww}{{\overline\Sigma}_{_{WW}}}
\newcommand{\bStg}{{\overline\Sigma}_{_{3Q}}}
\newcommand{\bStt}{{\overline\Sigma}_{_{33}}}
\newcommand{\Sssn}{\Sigma_{_{\hkn\hkn}}}
\newcommand{\Sssc}{\Sigma_{_{\phi\phi}}}
\newcommand{\Szn}{\Sigma_{_{\zb\hkn}}}
\newcommand{\Swc}{\Sigma_{_{\wb\hkg}}}
\newcommand{\mix}[2]{{\cal{M}}^{#1}\lpar{#2}\rpar}
\newcommand{\bmix}[2]{\Pi^{{#1},\ssF}_{_{\zb\ab}}\lpar{#2}\rpar}
\newcommand{\hPgg}[2]{{\hat{\Pi}^{{#1},\ssF}}_{\ph\ph}\lpar{#2}\rpar}
\newcommand{\hmix}[2]{{\hat{\Pi}^{{#1},\ssF}}_{_{\zb\ab}}\lpar{#2}\rpar}
\newcommand{\Dz}[2]{{\cal{D}}_{_{\zb}}^{#1}\lpar{#2}\rpar}
\newcommand{\bDz}[2]{{\cal{D}}^{{#1},\ssF}_{_{\zb}}\lpar{#2}\rpar}
\newcommand{\hDz}[2]{{\hat{\cal{D}}}^{{#1},\ssF}_{_{\zb}}\lpar{#2}\rpar}
\newcommand{\Szzd}[2]{\Sigma'^{#1}_{_{\zb\zb}}\lpar{#2}\rpar}
\newcommand{\Swwd}[2]{\Sigma'^{#1}_{_{\wb\wb}}\lpar{#2}\rpar}
\newcommand{\Shhd}[2]{\Sigma'^{#1}_{_{\hb\hb}}\lpar{#2}\rpar}
\newcommand{\ZFren}[2]{{\cal{Z}}^{#1}\lpar{#2}\rpar}
\newcommand{\WFren}[2]{{\cal{W}}^{#1}\lpar{#2}\rpar}
\newcommand{\HFren}[2]{{\cal{H}}^{#1}\lpar{#2}\rpar}
\newcommand{\WI}{\cal{W}}
%
%
\newcommand{\cf}{c_f}
\newcommand{\Cf}{C_{_F}}
\newcommand{\Nf}{N_f}
\newcommand{\Nc}{N_c}
\newcommand{\Ncs}{N^2_c}
\newcommand{\nf }{n_f}
\newcommand{\nfs}{n^2_f}
\newcommand{\nfc}{n^3_f}
\newcommand{\MSB}{\overline{MS}}
\newcommand{\LMSB}{\Lambda_{\overline{\mathrm{MS}}}}
\newcommand{\LMSBp}{\Lambda'_{\overline{\mathrm{MS}}}}
\newcommand{\LMSBS}{\Lambda^2_{\overline{\mathrm{MS}}}}
\newcommand{\LMSBv }{\mbox{$\Lambda^{(5)}_{\overline{\mathrm{MS}}}$}}
\newcommand{\LMSBvS}{\mbox{$\left(\Lambda^{(5)}_{\overline{\mathrm{MS}}}\right)^2$}}
\newcommand{\LMSBt }{\mbox{$\Lambda^{(3)}_{\overline{\mathrm{MS}}}$}}
\newcommand{\LMSBtS}{\mbox{$\left(\Lambda^{(3)}_{\overline{\mathrm{MS}}}\right)^2$}}
\newcommand{\LMSBf }{\mbox{$\Lambda^{(4)}_{\overline{\mathrm{MS}}}$}}
\newcommand{\LMSBfS}{\mbox{$\left(\Lambda^{(4)}_{\overline{\mathrm{MS}}}\right)^2$}}
\newcommand{\LMSBn }{\mbox{$\Lambda^{(\nf)}_{\overline{\mathrm{MS}}}$}}
\newcommand{\LMSBnS}{\mbox{$\left(\Lambda^{(\nf)}_{\overline{\mathrm{MS}}}\right)^2$}}
\newcommand{\LMSBnml }{\mbox{$\Lambda^{(\nf-1)}_{\overline{\mathrm{MS}}}$}}
\newcommand{\LMSBnmlS}{\mbox{$\left(\Lambda^{(\nf-1)}_{\overline{\mathrm{MS}}}\right)^2$}}
\newcommand{\Bnf}{\lpar\nf \rpar}
\newcommand{\Bnfm}{\lpar\nf-1 \rpar}
\newcommand{\LuM}{L_{_M}}
\newcommand{\bef}{\beta_{\ff}}
\newcommand{\befs}{\beta^2_{\ff}}
\newcommand{\befc}{\beta^3_{f}}
\newcommand{\alsp}{\alpha'_{_S}}
\newcommand{\api}{\displaystyle \frac{\als(s)}{\pi}}
\newcommand{\alss}{\alpha^2_{_S}}
\newcommand{\ztwo}{\zeta(2)}
\newcommand{\ztri}{\zeta(3)}
\newcommand{\zfor}{\zeta(4)}
\newcommand{\zfiv}{\zeta(5)}
\newcommand{\bi}[1]{b_{#1}}
\newcommand{\ci}[1]{c_{#1}}
\newcommand{\Ci}[1]{C_{#1}}
\newcommand{\bip}[1]{b'_{#1}}
\newcommand{\cip}[1]{c'_{#1}}
%
%
\newcommand{\osps}{16\,\pi^2}
\newcommand{\srt}{\sqrt{2}}
\newcommand{\ospsi}{\displaystyle{\frac{i}{16\,\pi^2}}}
%
%
\newcommand{\tfpromu}{\mbox{$e^+e^-\to \mu^+\mu^-$}}
\newcommand{\tfprotau}{\mbox{$e^+e^-\to \tau^+\tau^-$}}
\newcommand{\tfproe}{\mbox{$e^+e^-\to e^+e^-$}}
\newcommand{\tfpronu}{\mbox{$e^+e^-\to \barnu\nu$}}
\newcommand{\tfproqq}{\mbox{$e^+e^-\to \barq q$}}
\newcommand{\tfprohad}{\mbox{$e^+e^-\to\,$} hadrons}
%
%
\newcommand{\bpromu}{\mbox{$e^+e^-\to \mu^+\mu^-\ph$}}
\newcommand{\bprotau}{\mbox{$e^+e^-\to \tau^+\tau^-\ph$}}
\newcommand{\bproe}{\mbox{$e^+e^-\to e^+e^-\ph$}}
\newcommand{\bpronu}{\mbox{$e^+e^-\to \barnu\nu\ph$}}
\newcommand{\bproqq}{\mbox{$e^+e^-\to \barq q \ph$}}
%
%
\newcommand{\tbprow} {\mbox{$e^+e^-\to \wbp \wbm $}}
\newcommand{\tbproz} {\mbox{$e^+e^-\to \zb  \zb  $}}
\newcommand{\tbproh} {\mbox{$e^+e^-\to \zb  \hb  $}}
\newcommand{\tbprozg}{\mbox{$e^+e^-\to \zb  \ph  $}}
\newcommand{\tbprog} {\mbox{$e^+e^-\to \ph  \ph  $}}
%
%
\newcommand{\Fermionline}[1]{
\vcenter{\hbox{
  \begin{picture}(60,20)(0,{#1})
  \SetScale{2.}
    \ArrowLine(0,5)(30,5)
  \end{picture}}}
}
\newcommand{\AntiFermionline}[1]{
\vcenter{\hbox{
  \begin{picture}(60,20)(0,{#1})
  \SetScale{2.}
    \ArrowLine(30,5)(0,5)
  \end{picture}}}
}
\newcommand{\Photonline}[1]{
\vcenter{\hbox{
  \begin{picture}(60,20)(0,{#1})
  \SetScale{2.}
    \Photon(0,5)(30,5){2}{6.5}
  \end{picture}}}
}
\newcommand{\Gluonline}[1]{
\vcenter{\hbox{
  \begin{picture}(60,20)(0,{#1})
  \SetScale{2.}
    \Gluon(0,5)(30,5){2}{6.5}
  \end{picture}}}
}
\newcommand{\Wbosline}[1]{
\vcenter{\hbox{
  \begin{picture}(60,20)(0,{#1})
  \SetScale{2.}
    \Photon(0,5)(30,5){2}{4}
    \ArrowLine(13.3,3.1)(16.9,7.2)
  \end{picture}}}
}
\newcommand{\Zbosline}[1]{
\vcenter{\hbox{
  \begin{picture}(60,20)(0,{#1})
  \SetScale{2.}
    \Photon(0,5)(30,5){2}{4}
  \end{picture}}}
}
\newcommand{\Philine}[1]{
\vcenter{\hbox{
  \begin{picture}(60,20)(0,{#1})
  \SetScale{2.}
    \DashLine(0,5)(30,5){2}
  \end{picture}}}
}
\newcommand{\Phicline}[1]{
\vcenter{\hbox{
  \begin{picture}(60,20)(0,{#1})
  \SetScale{2.}
    \DashLine(0,5)(30,5){2}
    \ArrowLine(14,5)(16,5)
  \end{picture}}}
}
\newcommand{\Ghostline}[1]{
\vcenter{\hbox{
  \begin{picture}(60,20)(0,{#1})
  \SetScale{2.}
    \DashLine(0,5)(30,5){.5}
    \ArrowLine(14,5)(16,5)
  \end{picture}}}
}
%
%
\newcommand{\gauge}{g}
\newcommand{\gpar}{\xi}
\newcommand{\gparA}{\xi_{_A}}
\newcommand{\gparZ}{\xi_{_Z}}
\newcommand{\gpari}[1]{\gpar_{#1}}
\newcommand{\gparis}[1]{\gpar^2_{#1}}
\newcommand{\gpariq}[1]{\gpar^4_{#1}}
\newcommand{\gpars}{\xi^2}
\newcommand{\dgpar}{\Delta\gpar}
\newcommand{\dgparA}{\Delta\gparA}
\newcommand{\dgparZ}{\Delta\gparZ}
\newcommand{\gparq}{\xi^4}
\newcommand{\gparAs}{\xi^2_{_A}}
\newcommand{\gparAq}{\xi^4_{_A}}
\newcommand{\gparZs}{\xi^2_{_Z}}
\newcommand{\gparZq}{\xi^4_{_Z}}
\newcommand{\Rxi}{R_{\gpar}}
\newcommand{\UG}{U}
\newcommand{\UGi}{\ssU}
\newcommand{\hxi}{\chi}
%
%
\newcommand{\LSM}{{\cal{L}}_{_{\rm{SM}}}}
\newcommand{\LSMr}{{\cal{L}}^{\rm{\ssR}}_{_{\rm{SM}}}}
\newcommand{\LYM}{{\cal{L}}_{_{\rm{YM}}}}
\newcommand{\Lzer}{{\cal{L}}_{0}}
\newcommand{\Lone}{{\cal{L}}^{{\bos},{\rm{I}}}}
\newcommand{\Lpro}{{\cal{L}}_{\rm{prop}}}
\newcommand{\Ls  }{{\cal{L}}_{_{\rm{S}}}}
\newcommand{\Lsi }{{\cal{L}}^{\rm{I}}_{_{\rm{S}}}}
\newcommand{\Lgf }{{\cal{L}}_{\rm{gf}}}
\newcommand{\Lgfi}{{\cal{L}}^{\rm{I}}_{\rm{gf}}}
\newcommand{\Lf  }{{\cal{L}}^{{\fer},{\rm{I}}}_{\ssV}}
\newcommand{\LHf }{{\cal{L}}^{\fer}_{\ssS}}
\newcommand{\LHfm}{{\cal{L}}^{{\fer},m}_{\ssS}}
\newcommand{\LHfi}{{\cal{L}}^{{\fer},{\rm{I}}}_{\ssS}}
\newcommand{\Lren}{{\cal{L}}_{\rm{\ssR}}}
\newcommand{\Lct}{{\cal{L}}_{\rm{ct}}}
\newcommand{\Lcti}[1]{{\cal{L}}^{#1}_{\rm{ct}}}
\newcommand{\LctI}{{\cal{L}}^{(2)}_{\rm{ct}}}
\newcommand{\Llone}{{\cal{L}}}
\newcommand{\LQED}{{\cal{L}}_{_{\rm{QED}}}}
\newcommand{\LQEDz}{{\cal{L}}^{0}_{_{\rm{QED}}}}
\newcommand{\LQEDi}{{\cal{L}}^{\rm{\ssI}}_{_{\rm{QED}}}}
\newcommand{\LQEDr}{{\cal{L}}^{\rm{\ssR}}_{_{\rm{QED}}}}
\newcommand{\Greenf}{G}
\newcommand{\Greenfa}[1]{G\lpar{#1}\rpar}
\newcommand{\Greenft}[2]{G\lpar{#1,#2}\rpar}
\newcommand{\FST}[3]{F_{#1#2}^{#3}}
\newcommand{\cD}[1]{D_{#1}}
\newcommand{\pd}[1]{\partial_{#1}}
\newcommand{\tgen}[1]{\tau^{#1}}
\newcommand{\gbl}{g_1}
\newcommand{\lctt}[3]{\varepsilon_{#1#2#3}}
\newcommand{\lctf}[4]{\varepsilon_{#1#2#3#4}}
\newcommand{\lctfb}[4]{\varepsilon\lpar{#1#2#3#4}\rpar}
\newcommand{\slct}{\varepsilon}
\newcommand{\cgfi}[1]{{\cal{C}}^{#1}}
\newcommand{\cgfZ}{{\cal{C}}^{\ssZ}}
\newcommand{\cgfA}{{\cal{C}}^{\ssA}}
\newcommand{\hpms}{\mu^2}
\newcommand{\hpal}{\alpha_{_H}}
\newcommand{\hpals}{\alpha^2_{_H}}
\newcommand{\hpbe}{\beta_{_H}}
\newcommand{\hpbep}{\beta^{'}_{_H}}
\newcommand{\hpla}{\lambda}
\newcommand{\hpalf}{\alpha_{f}}
\newcommand{\hpbef}{\beta_{f}}
\newcommand{\tpar}[1]{\Lambda^{#1}}
\newcommand{\Mop}[2]{{\rm{M}}^{#1#2}}
\newcommand{\Lop}[2]{{\rm{L}}^{#1#2}}
\newcommand{\Lgen}[1]{T^{#1}}
\newcommand{\Rgen}[1]{t^{#1}}
\newcommand{\fpari}[1]{\lambda_{#1}}
\newcommand{\fQ}[1]{Q_{#1}}
\newcommand{\unm}{I}
\newcommand{\cDsla}{/\!\!\!\!D}
%
%
\newcommand{\saff}[1]{A_{#1}}                    
\newcommand{\aff}[2]{A_{#1}\lpar #2\rpar}
\newcommand{\sbff}[1]{B_{#1}}                    
\newcommand{\sfbff}[1]{B^{\ssF}_{#1}}
\newcommand{\bff}[4]{B_{#1}\lpar #2;#3,#4\rpar}
\newcommand{\bfft}[3]{B_{#1}\lpar #2,#3\rpar}
\newcommand{\fbff}[4]{B^{\ssF}_{#1}\lpar #2;#3,#4\rpar}
\newcommand{\cdbff}[4]{\Delta B_{#1}\lpar #2;#3,#4\rpar}
\newcommand{\sdbff}[4]{\delta B_{#1}\lpar #2;#3,#4\rpar}
\newcommand{\cdbfft}[3]{\Delta B_{#1}\lpar #2,#3\rpar}
\newcommand{\sdbfft}[3]{\delta B_{#1}\lpar #2,#3\rpar}
\newcommand{\scff}[1]{C_{#1}}                    
\newcommand{\scffo}[2]{C_{#1}\lpar{#2}\rpar}
\newcommand{\cff}[7]{C_{#1}\lpar #2,#3,#4;#5,#6,#7\rpar}
\newcommand{\sccff}[5]{c_{#1}\lpar #2;#3,#4,#5\rpar}
\newcommand{\sdff}[1]{D_{#1}}                    
\newcommand{\dffp}[7]{D_{#1}\lpar #2,#3,#4,#5,#6,#7;}
\newcommand{\dffm}[4]{#1,#2,#3,#4\rpar}
\newcommand{\bzfa}[2]{B^{\ssF}_{_{#2}}\lpar{#1}\rpar}
\newcommand{\bzfaa}[3]{B^{\ssF}_{_{#2#3}}\lpar{#1}\rpar}
\newcommand{\shcff}[4]{C_{_{#2#3#4}}\lpar{#1}\rpar}
\newcommand{\shdff}[6]{D_{_{#3#4#5#6}}\lpar{#1,#2}\rpar}
\newcommand{\scdff}[3]{d_{#1}\lpar #2,#3\rpar}
\newcommand{\scaldff}[1]{{\cal{D}}^{#1}}
\newcommand{\caldff}[2]{{\cal{D}}^{#1}\lpar{#2}\rpar}
\newcommand{\caldfft}[3]{{\cal{D}}_{#1}^{#2}\lpar{#3}\rpar}
%
%
\newcommand{\slaff}[1]{a_{#1}}
\newcommand{\slbff}[1]{b_{#1}}
\newcommand{\slbffh}[1]{{\hat{b}}_{#1}}
\newcommand{\ssldff}[1]{d_{#1}}
\newcommand{\sslcff}[1]{c_{#1}}
\newcommand{\slcff}[2]{c_{#1}^{(#2)}}
\newcommand{\sldff}[2]{d_{#1}^{(#2)}}
\newcommand{\lbff}[3]{b_{#1}\lpar #2;#3\rpar}
\newcommand{\lbffh}[2]{{\hat{b}}_{#1}\lpar #2\rpar}
\newcommand{\lcff}[8]{c_{#1}^{(#2)}\lpar  #3,#4,#5;#6,#7,#8\rpar}
\newcommand{\ldffp}[8]{d_{#1}^{(#2)}\lpar #3,#4,#5,#6,#7,#8;}
\newcommand{\ldffm}[4]{#1,#2,#3,#4\rpar}
%
%
\newcommand{\Iff}[4]{I_{#1}\lpar #2;#3,#4 \rpar}
\newcommand{\Jff}[4]{J_{#1}\lpar #2;#3,#4 \rpar}
\newcommand{\Jds}[5]{{\bar{J}}_{#1}\lpar #2,#3;#4,#5 \rpar}
\newcommand{\sJds}[1]{{\bar{J}}_{#1}}
%
\newcommand{\nhmt}{\frac{n}{2}-2}
\newcommand{\nhmts}{{n}/{2}-2}
\newcommand{\omnh}{1-\frac{n}{2}}
\newcommand{\nhmo}{\frac{n}{2}-1}
\newcommand{\fmon}{4-n}
\newcommand{\lpi}{\ln\pi}
\newcommand{\lmass}[1]{\ln #1}
\newcommand{\egnh}{\egam{\frac{n}{2}}}
\newcommand{\egomnh}{\egam{1-\frac{n}{2}}}
\newcommand{\egtmnh}{\egam{2-\frac{n}{2}}}
\newcommand{\Ddr}{{\ds\frac{1}{{\bar{\varepsilon}}}}}
\newcommand{\Ddrs}{{\ds\frac{1}{{\bar{\varepsilon}^2}}}}
\newcommand{\Ddrd}{{\bar{\varepsilon}}}
\newcommand{\ept}{\hat\varepsilon}
\newcommand{\Ddrh}{{\ds\frac{1}{\hat{\varepsilon}}}}
\newcommand{\Ddrp}{{\ds\frac{1}{\varepsilon'}}}
\newcommand{\Ddrps}{\lpar{\ds{\frac{1}{\varepsilon'}}}\rpar^2}
\newcommand{\dre}{\varepsilon}
\newcommand{\drei}[1]{\varepsilon_{#1}}
\newcommand{\epp}{\varepsilon'}
\newcommand{\eps}{\varepsilon^*}
\newcommand{\ep}{\epsilon}
\newcommand{\propbt}[6]{{{#1_{#2}#1_{#3}}\over{\lpar #1^2 + #4
-\ib\ep\rpar\lpar\lpar #5\rpar^2 + #6 -\ib\ep\rpar}}}
\newcommand{\propbo}[5]{{{#1_{#2}}\over{\lpar #1^2 + #3 - \ib\ep\rpar
\lpar\lpar #4\rpar^2 + #5 -\ib\ep\rpar}}}
\newcommand{\propc}[6]{{1\over{\lpar #1^2 + #2 - \ib\ep\rpar
\lpar\lpar #3\rpar^2 + #4 -\ib\ep\rpar
\lpar\lpar #5\rpar^2 + #6 -\ib\ep\rpar}}}
\newcommand{\propa}[2]{{1\over {#1^2 + #2^2 - \ib\ep}}}
\newcommand{\propb}[4]{{1\over {\lpar #1^2 + #2 - \ib\ep\rpar
\lpar\lpar #3\rpar^2 + #4 -\ib\ep\rpar}}}
\newcommand{\propbs}[4]{{1\over {\lpar\lpar #1\rpar^2 + #2 - \ib\ep\rpar
\lpar\lpar #3\rpar^2 + #4 -\ib\ep\rpar}}}
\newcommand{\propat}[4]{{#3_{#1}#3_{#2}\over {#3^2 + #4^2 - \ib\ep}}}
\newcommand{\propaf}[6]{{#5_{#1}#5_{#2}#5_{#3}#5_{#4}\over
{#5^2 + #6^2 -\ib\ep}}}
\newcommand{\momeps}[1]{#1^2 - \ib\ep}
\newcommand{\mopeps}[1]{#1^2 + \ib\ep}
\newcommand{\propz}[1]{{1\over{#1^2 + \mzs - \ib\ep}}}
\newcommand{\propw}[1]{{1\over{#1^2 + \mws - \ib\ep}}}
\newcommand{\proph}[1]{{1\over{#1^2 + \mhs - \ib\ep}}}
\newcommand{\propf}[2]{{1\over{#1^2 + #2}}}
\newcommand{\propzrg}[3]{{{\delta_{#1#2}}\over{{#3}^2 + \mzs - \ib\ep}}}
\newcommand{\propwrg}[3]{{{\delta_{#1#2}}\over{{#3}^2 + \mws - \ib\ep}}}
\newcommand{\propzug}[3]{{
      {\delta_{#1#2} + \displaystyle{{{#3}^{#1}{#3}^{#2}}\over{\mzs}}}
                         \over{{#3}^2 + \mzs - \ib\ep}}}
\newcommand{\propwug}[3]{{
      {\delta_{#1#2} + \displaystyle{{{#3}^{#1}{#3}^{#2}}\over{\mws}}}
                        \over{{#3}^2 + \mws - \ib\ep}}}
\newcommand{\thf}[1]{\theta\lpar #1\rpar}
\newcommand{\epf}[1]{\varepsilon\lpar #1\rpar}
\newcommand{\singp}{\stackrel{\rm{sing}}{\rightarrow}}
\newcommand{\aint}[3]{\int_{#1}^{#2}\,d #3}
\newcommand{\aroot}[1]{\sqrt{#1}}
\newcommand{\gramc}{\Delta_3}
\newcommand{\gramd}{\Delta_4}
\newcommand{\pinch}[2]{P^{(#1)}\lpar #2\rpar}
\newcommand{\pinchc}[2]{C^{(#1)}_{#2}}
\newcommand{\pinchd}[2]{D^{(#1)}_{#2}}
\newcommand{\loarg}[1]{\ln\lpar #1\rpar}
\newcommand{\loargr}[1]{\ln\lrbr #1\rrbr}
\newcommand{\lsoarg}[1]{\ln^2\lpar #1\rpar}
\newcommand{\ltarg}[2]{\ln\lpar #1\rpar\lpar #2\rpar}
\newcommand{\rfun}[2]{R\lpar #1,#2\rpar}
\newcommand{\pinchb}[3]{B_{#1}\lpar #2,#3\rpar}
\newcommand{\lga}{\ph}
\newcommand{\lzga}{\ssZ\ph}
%
%
\newcommand{\afa}[5]{A_{#1}^{#2}\lpar #3;#4,#5\rpar}
\newcommand{\bfa}[5]{B_{#1}^{#2}\lpar #3;#4,#5\rpar}
\newcommand{\hfa}[5]{H_{#1}^{#2}\lpar #3;#4,#5\rpar}
\newcommand{\rfa}[5]{R_{#1}^{#2}\lpar #3;#4,#5\rpar}
\newcommand{\afao}[3]{A_{#1}^{#2}\lpar #3\rpar}
\newcommand{\bfao}[3]{B_{#1}^{#2}\lpar #3\rpar}
\newcommand{\hfao}[3]{H_{#1}^{#2}\lpar #3\rpar}
\newcommand{\rfao}[3]{R_{#1}^{#2}\lpar #3\rpar}
\newcommand{\afax}[6]{A_{#1}^{#2}\lpar #3;#4,#5,#6\rpar}
\newcommand{\afas}[2]{A_{#1}^{#2}}
\newcommand{\bfas}[2]{B_{#1}^{#2}}
\newcommand{\hfas}[2]{H_{#1}^{#2}}
\newcommand{\rfas}[2]{R_{#1}^{#2}}
\newcommand{\tfas}[2]{T_{#1}^{#2}}
\newcommand{\afaR}[6]{A_{#1}^{\gpar}\lpar #2;#3,#4,#5,#6 \rpar}
\newcommand{\bfaR}[6]{B_{#1}^{\gpar}\lpar #2;#3,#4,#5,#6 \rpar}
\newcommand{\hfaR}[6]{H_{#1}^{\gpar}\lpar #2;#3,#4,#5,#6 \rpar}
\newcommand{\shfaR}[1]{H_{#1}^{\gpar}}
\newcommand{\rfaR}[6]{R_{#1}^{\gpar}\lpar #2;#3,#4,#5,#6 \rpar}
\newcommand{\srfaR}[1]{R_{#1}^{\gpar}}
\newcommand{\afaRg}[5]{A_{#1 \lga}^{\gpar}\lpar #2;#3,#4,#5 \rpar}
\newcommand{\bfaRg}[5]{B_{#1 \lga}^{\gpar}\lpar #2;#3,#4,#5 \rpar}
\newcommand{\hfaRg}[5]{H_{#1 \lga}^{\gpar}\lpar #2;#3,#4,#5 \rpar}
\newcommand{\shfaRg}[1]{H_{#1\lga}^{\gpar}}
\newcommand{\rfaRg}[5]{R_{#1 \lga}^{\gpar}\lpar #2;#3,#4,#5 \rpar}
\newcommand{\srfaRg}[1]{R_{#1\lga}^{\gpar}}
\newcommand{\afaRt}[3]{A_{#1}^{\gpar}\lpar #2,#3 \rpar}
\newcommand{\hfaRt}[3]{H_{#1}^{\gpar}\lpar #2,#3 \rpar}
\newcommand{\hfaRf}[4]{H_{#1}^{\gpar}\lpar #2,#3,#4 \rpar}
\newcommand{\afasm}[4]{A_{#1}^{\lpar #2,#3,#4 \rpar}}
\newcommand{\bfasm}[4]{B_{#1}^{\lpar #2,#3,#4 \rpar}}
\newcommand{\htf}[2]{H_2\lpar #1,#2\rpar}
\newcommand{\rof}[2]{R_1\lpar #1,#2\rpar}
\newcommand{\rtf}[2]{R_3\lpar #1,#2\rpar}
\newcommand{\rtrans}[2]{R_{#1}^{#2}}
\newcommand{\momf}[2]{#1^2_{#2}}
\newcommand{\Scalvert}[8][70]{
  \vcenter{\hbox{
  \SetScale{0.8}
  \begin{picture}(#1,50)(15,15)
    \Line(25,25)(50,50)      \Text(34,20)[lc]{#6} \Text(11,20)[lc]{#3}
    \Line(50,50)(25,75)      \Text(34,60)[lc]{#7} \Text(11,60)[lc]{#4}
    \Line(50,50)(90,50)      \Text(11,40)[lc]{#2} \Text(55,33)[lc]{#8}
    \GCirc(50,50){10}{1}          \Text(60,48)[lc]{#5}
  \end{picture}}}
  }
%
%
\newcommand{\tHs}{\mu}
\newcommand{\tHsz}{\mu_{_0}}
\newcommand{\tHss}{\mu^2}
\newcommand{\Reb}{{\rm{Re}}}
\newcommand{\Imb}{{\rm{Im}}}
%
%
\newcommand{\spd}{\partial}
\newcommand{\ffun}[2]{F_{#1}\lpar #2\rpar}
\newcommand{\gfun}[2]{G_{#1}\lpar #2\rpar}
\newcommand{\sffun}[1]{F_{#1}}
\newcommand{\csffun}[1]{{\cal{F}}_{#1}}
\newcommand{\sgfun}[1]{G_{#1}}
\newcommand{\tpfi}{\lpar 2\pi\rpar^4\ib}
\newcommand{\ffv}{F_{_V}}
\newcommand{\fga}{G_{_A}}
\newcommand{\ffm}{F_{_M}}
\newcommand{\ffs}{F_{_S}}
\newcommand{\fgp}{G_{_P}}
\newcommand{\fge}{G_{_E}}
\newcommand{\ffa}{F_{_A}}
\newcommand{\ffps}{F_{_P}}
\newcommand{\ffe}{F_{_E}}
\newcommand{\gacom}[2]{\lpar #1 + #2\gfd\rpar}
\newcommand{\mft}{m_{\tilde f}}
\newcommand{\qft}{Q_{f'}}
\newcommand{\vft}{v_{\tilde f}}
\newcommand{\subb}[2]{b_{#1}\lpar #2 \rpar}
\newcommand{\fwfr}[5]{\Sigma\lpar #1,#2,#3;#4,#5 \rpar}
\newcommand{\slim}[2]{\lim_{#1 \to #2}}
\newcommand{\sprop}[3]{
{#1\over {\lpar q^2\rpar^2\lpar \lpar q+ #2\rpar^2+#3^2\rpar }}}
%
%
\newcommand{\xroot}[1]{x_{#1}}
\newcommand{\yroot}[1]{y_{#1}}
\newcommand{\zroot}[1]{z_{#1}}
\newcommand{\lvar}{l}
\newcommand{\rvar}{r}
\newcommand{\tvar}{t}
\newcommand{\uvar}{u}
\newcommand{\vvar}{v}
\newcommand{\xvar}{x}
\newcommand{\yvar}{y}
\newcommand{\zvar}{z}
\newcommand{\yvarp}{y'}
\newcommand{\rvars}{r^2}
\newcommand{\vvars}{v^2}
\newcommand{\xvars}{x^2}
\newcommand{\yvars}{y^2}
\newcommand{\zvars}{z^2}
\newcommand{\rvarc}{r^3}
\newcommand{\xvarc}{x^3}
\newcommand{\yvarc}{y^3}
\newcommand{\zvarc}{z^3}
\newcommand{\rvarq}{r^4}
\newcommand{\xvarq}{x^4}
\newcommand{\yvarq}{y^4}
\newcommand{\zvarq}{z^4}
\newcommand{\avar}{a}
\newcommand{\avars}{a^2}
\newcommand{\avarc}{a^3}
\newcommand{\avari}[1]{a_{#1}}
\newcommand{\avart}[2]{a_{#1}^{#2}}
\newcommand{\delvari}[1]{\delta_{#1}}
\newcommand{\rvari}[1]{r_{#1}}
\newcommand{\xvari}[1]{x_{#1}}
\newcommand{\yvari}[1]{y_{#1}}
\newcommand{\zvari}[1]{z_{#1}}
\newcommand{\rvart}[2]{r_{#1}^{#2}}
\newcommand{\xvart}[2]{x_{#1}^{#2}}
\newcommand{\yvart}[2]{y_{#1}^{#2}}
\newcommand{\zvart}[2]{z_{#1}^{#2}}
\newcommand{\rvaris}[1]{r^2_{#1}}
\newcommand{\xvaris}[1]{x^2_{#1}}
\newcommand{\yvaris}[1]{y^2_{#1}}
\newcommand{\zvaris}[1]{z^2_{#1}}
\newcommand{\Xvar}{X}
\newcommand{\Xvars}{X^2}
\newcommand{\Xvari}[1]{X_{#1}}
\newcommand{\Xvaris}[1]{X^2_{#1}}
\newcommand{\Yvar}{Y}
\newcommand{\Yvars}{Y^2}
\newcommand{\Yvari}[1]{Y_{#1}}
\newcommand{\Yvaris}[1]{Y^2_{#1}}
\newcommand{\lnx}{\ln\xvar}
\newcommand{\lnz}{\ln\zvar}
\newcommand{\lnsx}{\ln^2\xvar}
\newcommand{\lnsz}{\ln^2\zvar}
\newcommand{\lncz}{\ln^3\zvar}
\newcommand{\lnomz}{\ln\lpar 1-\zvar\rpar}
\newcommand{\lnsomz}{\ln^2\lpar 1-\zvar\rpar}
\newcommand{\ccoefi}[1]{c_{#1}}
\newcommand{\ccoeft}[2]{c^{#1}_{#2}}
%
%
\newcommand{\Smat}{{\cal{S}}}
\newcommand{\Mmat}{{\cal{M}}}
\newcommand{\Rmat}{{\cal{R}}}
\newcommand{\Xmat}[1]{X_{#1}}
\newcommand{\XmatI}[1]{X^{-1}_{#1}}
\newcommand{\unitmat}{I}
\newcommand{\zeromat}{O}
\newcommand{\paulimat}[1]{\tau_{#1}}
\newcommand{\Umat  }{U}
\newcommand{\Umath }{U^{+}}
\newcommand{\UmatL }{{\cal{U}}_{\ssL}}
\newcommand{\UmatLh}{{\cal{U}}^{+}_{\ssL}}
\newcommand{\UmatR }{{\cal{U}}_{\ssR}}
\newcommand{\UmatRh}{{\cal{U}}^{+}_{\ssR}}
\newcommand{\Vmat  }{V}
\newcommand{\Vmath }{V^{+}}
\newcommand{\VmatL }{{\cal{D}}_{\ssL}}
\newcommand{\VmatLh}{{\cal{D}}^{+}_{\ssL}}
\newcommand{\VmatR }{{\cal{D}}_{\ssR}}
\newcommand{\VmatRh}{{\cal{D}}^{+}_{\ssR}}
\newcommand{\Kmat}{{C}}
\newcommand{\Kmatc}{{C}^{\dagger}}
\newcommand{\Kmati}[1]{{C}_{#1}}
\newcommand{\Kmatci}[1]{{C}^{\dagger}_{#1}}
\newcommand{\ffac}[2]{f_{#1}^{#2}}
\newcommand{\Ffac}[1]{F_{#1}}
\newcommand{\Rvec}[2]{R^{(#1)}_{#2}}
\newcommand{\momfl}[2]{#1_{#2}}
\newcommand{\momfs}[2]{#1^2_{#2}}
\newcommand{\fpseZ}{A^{\ssF\ssP,\ssZ}}
\newcommand{\fpseA}{A^{\ssF\ssP,\ssA}}
\newcommand{\fptZ}{T^{\ssF\ssP,\ssZ}}
\newcommand{\fptA}{T^{\ssF\ssP,\ssA}}
\newcommand{\dprop}{\overline\Delta}
\newcommand{\dpropi}[1]{d_{#1}}
\newcommand{\dpropic}[1]{d^{c}_{#1}}
\newcommand{\dpropii}[2]{d_{#1}\lpar #2\rpar}
\newcommand{\dpropis}[1]{d^2_{#1}}
\newcommand{\dproppi}[1]{d'_{#1}}
\newcommand{\psf}[4]{P\lpar #1;#2,#3,#4\rpar}
\newcommand{\ssf}[5]{S^{(#1)}\lpar #2;#3,#4,#5\rpar}
\newcommand{\csf}[5]{C_{_S}^{(#1)}\lpar #2;#3,#4,#5\rpar}
%
%
\newcommand{\lvec}{l}
\newcommand{\lvecs}{l^2}
\newcommand{\lveci}[1]{l_{#1}}
\newcommand{\mvec}{m}
\newcommand{\mvecs}{m^2}
\newcommand{\mveci}[1]{m_{#1}}
\newcommand{\nvec}{n}
\newcommand{\nvecs}{n^2}
\newcommand{\nveci}[1]{n_{#1}}
\newcommand{\epi}[1]{\epsilon_{#1}}
\newcommand{\phep}[1]{\ep_{#1}}
\newcommand{\php}[3]{\ep^{#1}_{#2}\lpar #3 \rpar}
\newcommand{\sphep}{\ep}
\newcommand{\vbep}[1]{e_{#1}}
\newcommand{\vbepp}[1]{e^{+}_{#1}}
\newcommand{\vbepm}[1]{e^{-}_{#1}}
\newcommand{\svbep}{e}
%
%
\newcommand{\lpol}{\lambda}
\newcommand{\spol}{\sigma}
\newcommand{\rpol}{\rho  }
\newcommand{\kpol}{\kappa}
\newcommand{\lpols}{\lambda^2}
\newcommand{\spols}{\sigma^2}
\newcommand{\rpols}{\rho^2}
\newcommand{\kpols}{\kappa^2}
\newcommand{\lpoli}[1]{\lambda_{#1}}
\newcommand{\spoli}[1]{\sigma_{#1}}
\newcommand{\rpoli}[1]{\rho_{#1}}
\newcommand{\kpoli}[1]{\kappa_{#1}}
%
%
\newcommand{\uvec}{u}
\newcommand{\uveci}[1]{u_{#1}}
%
%
\newcommand{\imom}{q}
\newcommand{\imomi}[1]{q_{#1}}
\newcommand{\imoms}{q^2}
\newcommand{\pmom}{p}
\newcommand{\pmomp}{p'}
\newcommand{\pmoms}{p^2}
\newcommand{\pmomq}{p^4}
\newcommand{\pmomx}{p^6}
\newcommand{\pmomi}[1]{p_{#1}}
\newcommand{\pmompi}[1]{p'_{#1}}
\newcommand{\pmomis}[1]{p^2_{#1}}
\newcommand{\Pmom}{P}
\newcommand{\Pmoms}{P^2}
\newcommand{\Pmomi}[1]{P_{#1}}
\newcommand{\Pmomis}[1]{P^2_{#1}}
\newcommand{\Pmomp}{P'}
\newcommand{\Pmomps}{{P'}^2}
\newcommand{\Pmompi}[1]{P'_{#1}}
\newcommand{\Pmompis}[1]{{P'}^2_{#1}}
\newcommand{\Kmom}{K}
\newcommand{\Kmoms}{K^2}
\newcommand{\Kmomi}[1]{K_{#1}}
\newcommand{\Kmomis}[1]{K^2_{#1}}
\newcommand{\kmom}{k}
\newcommand{\kmoms}{k^2}
\newcommand{\kmomi}[1]{k_{#1}}
\newcommand{\lmom}{l}
\newcommand{\lmoms}{l^2}
\newcommand{\lmomi}[1]{l_{#1}}
\newcommand{\qmom}{q}
\newcommand{\qmoms}{q^2}
\newcommand{\qmomi}[1]{q_{#1}}
\newcommand{\qmomis}[1]{q^2_{#1}}
\newcommand{\smom}{s}
\newcommand{\smoms}{s^2}
\newcommand{\smomi}[1]{s_{#1}}
\newcommand{\tmom}{t}
\newcommand{\tmoms}{t^2}
\newcommand{\tmomi}[1]{t_{#1}}
\newcommand{\Trmom}{Q}
\newcommand{\Prmom}{P}
\newcommand{\gmv}{Q^2}
\newcommand{\Trmoms}{Q^2}
\newcommand{\Prmoms}{P^2}
\newcommand{\Ptmoms}{T^2}
\newcommand{\Pumoms}{U^2}
\newcommand{\Trmomq}{Q^4}
\newcommand{\Prmomq}{P^4}
\newcommand{\Ptmomq}{T^4}
\newcommand{\Pumomq}{U^4}
\newcommand{\Trmomx}{Q^6}
\newcommand{\Trmomi}[1]{Q_{#1}}
\newcommand{\Trmomis}[1]{Q^2_{#1}}
\newcommand{\Prmomi}[1]{P_{#1}}
\newcommand{\pone}{p_1}
\newcommand{\ptwo}{p_2}
\newcommand{\ptre}{p_3}
\newcommand{\pfor}{p_4}
\newcommand{\pones}{p_1^2}
\newcommand{\ptwos}{p_2^2}
\newcommand{\ptres}{p_3^2}
\newcommand{\pfors}{p_4^2}
\newcommand{\poneq}{p_1^4}
\newcommand{\ptwoq}{p_2^4}
\newcommand{\ptreq}{p_3^4}
\newcommand{\pforq}{p_4^4}
\newcommand{\modmom}[1]{\mid{\vec{#1}}\mid}
\newcommand{\modmoms}[1]{\mid{\vec{#1}}\mid^2}
\newcommand{\modmomi}[2]{\mid{\vec{#1}}_{#2}\mid}
\newcommand{\vect}[1]{{\vec{#1}}}
\newcommand{\Energ}{E}
\newcommand{\Energp}{E'}
\newcommand{\Energpp}{E''}
\newcommand{\Energs}{E^2}
\newcommand{\Energc}{E^3}
\newcommand{\Energf}{E^4}
\newcommand{\Energv}{E^5}
\newcommand{\Energx}{E^6}
\newcommand{\Energi}[1]{E_{#1}}
\newcommand{\Energt}[2]{E_{#1}^{#2}}
\newcommand{\Energis}[1]{E^2_{#1}}
\newcommand{\energ}{e}
\newcommand{\energp}{e'}
\newcommand{\energpp}{e''}
\newcommand{\energs}{e^2}
\newcommand{\energi}[1]{e_{#1}}
\newcommand{\energt}[2]{e_{#1}^{#2}}
\newcommand{\energis}[1]{e^2_{#1}}
\newcommand{\wenerg}{w}
\newcommand{\wenergs}{w^2}
\newcommand{\wenergi}[1]{w_{#1}}
\newcommand{\wenergp}{w'}
\newcommand{\wenergpp}{w''}
%
%
\newcommand{\ecut}{e}
\newcommand{\ecuts}{e^2}
\newcommand{\ecuti}[1]{e^{#1}}
\newcommand{\ccut}{c_m}
\newcommand{\ccuti}[1]{c_{#1}}
\newcommand{\ccuts}{c^2_m}
\newcommand{\scuts}{s^2_m}
\newcommand{\ccutis}[1]{c^2_{#1}}
\newcommand{\ccutic}[1]{c^3_{#1}}
\newcommand{\ccutc}{c^3_m}
\newcommand{\rcut}{\varrho}
\newcommand{\rcuts}{\varrho^2}
\newcommand{\rcuti}[1]{\varrho_{#1}}
\newcommand{\rcutu}[1]{\varrho^{#1}}
\newcommand{\Dcut}{\Delta}
%
\newcommand{\dwf}{\delta_{_{\rm{WF}}}}
\newcommand{\gbar}{\overline g}
\newcommand{\PP}{\mbox{PP}}
\newcommand{\mv}{m_{_V}}
\newcommand{\bGv}{{\overline\Gamma}_{_V}}
\newcommand{\Umuv}{\hat{\mu}_\ssV}
\newcommand{\Svv}{{\Sigma}_\ssV}
\newcommand{\muv}{p_\ssV}
\newcommand{\muvb}{\mu_{\ssV_{0}}}
\newcommand{\URPvv}{{P}_\ssV}
\newcommand{\RPvv}{{P}_\ssV}
\newcommand{\Svvrem}{{\Sigma}_\ssV^{\mathrm{rem}}}
\newcommand{\USvvrem}{\hat{\Sigma}_\ssV^{\mathrm{rem}}}
\newcommand{\Gv}{\Gamma_{_V}}
%
%
\newcommand{\param}{p}
\newcommand{\parami}[1]{p^{#1}}
\newcommand{\paramb}{p_{0}}
\newcommand{\Zcon}{Z}
\newcommand{\Zconi}[1]{Z_{#1}}
\newcommand{\zconi}[1]{z_{#1}}
\newcommand{\Zconim}[1]{{Z^-_{#1}}}
\newcommand{\zconim}[1]{{z^-_{#1}}}
\newcommand{\Zcont}[2]{Z_{#1}^{#2}}
\newcommand{\zcont}[2]{z_{#1}^{#2}}
\newcommand{\zcontm}[2]{z_{#1}^{{#2}-}}
\newcommand{\sZconi}[2]{\sqrt{Z_{#1}}^{\;#2}}
\newcommand{\gacome}[1]{\lpar #1 - \gfd\rpar}
\newcommand{\sPj}[2]{\Lambda^{#1}_{#2}}
\newcommand{\sPjs}[2]{\Lambda_{#1,#2}}
\newcommand{\amos}{\mbox{$M^2_{_1}$}}
\newcommand{\amts}{\mbox{$M^2_{_2}$}}
\newcommand{\er}{e_{_{R}}}
\newcommand{\epr}{e'_{_{R}}}
\newcommand{\ers}{e^2_{_{R}}}
\newcommand{\erc}{e^3_{_{R}}}
\newcommand{\erq}{e^4_{_{R}}}
\newcommand{\erf}{e^5_{_{R}}}
\newcommand{\sour}{J}
\newcommand{\sourb}{\overline J}
\newcommand{\lrm}{M_{_R}}
%
%
\newcommand{\vlami}[1]{\lambda_{#1}}
\newcommand{\vlamis}[1]{\lambda^2_{#1}}
\newcommand{\Vvert}{V}
\newcommand{\Avert}{A}
\newcommand{\Svert}{S}
\newcommand{\Pvert}{P}
\newcommand{\vvert}{F}
\newcommand{\Cvert}{\cal{V}}
\newcommand{\Bvert}{\cal{B}}
\newcommand{\Vveri}[2]{V_{#1}^{#2}}
\newcommand{\Fveri}[1]{{\cal{F}}^{#1}}
\newcommand{\Cveri}[1]{{\cal{V}}\lpar{#1}\rpar}
\newcommand{\Bveri}[1]{{\cal{B}}\lpar{#1}\rpar}
\newcommand{\Vverti}[3]{V_{#1}^{#2}\lpar{#3}\rpar}
\newcommand{\Averti}[3]{A_{#1}^{#2}\lpar{#3}\rpar}
\newcommand{\Gverti}[3]{G_{#1}^{#2}\lpar{#3}\rpar}
\newcommand{\Zverti}[3]{Z_{#1}^{#2}\lpar{#3}\rpar}
\newcommand{\Hverti}[2]{H^{#1}\lpar{#2}\rpar}
\newcommand{\Wverti}[3]{W_{#1}^{#2}\lpar{#3}\rpar}
\newcommand{\Cverti}[2]{{\cal{V}}_{#1}^{#2}}
\newcommand{\vverti}[3]{F^{#1}_{#2}\lpar{#3}\rpar}
\newcommand{\averti}[3]{{\overline{F}}^{#1}_{#2}\lpar{#3}\rpar}
\newcommand{\fveone}[1]{f_{#1}}
\newcommand{\fvetri}[3]{f^{#1}_{#2}\lpar{#3}\rpar}
\newcommand{\gvetri}[3]{g^{#1}_{#2}\lpar{#3}\rpar}
\newcommand{\cvetri}[3]{{\cal{F}}^{#1}_{#2}\lpar{#3}\rpar}
\newcommand{\hvetri}[3]{{\hat{\cal{F}}}^{#1}_{#2}\lpar{#3}\rpar}
\newcommand{\avetri}[3]{{\overline{\cal{F}}}^{#1}_{#2}\lpar{#3}\rpar}
\newcommand{\fverti}[2]{F^{#1}_{#2}}
\newcommand{\cverti}[2]{{\cal{F}}_{#1}^{#2}}
\newcommand{\fV}{f_{_{\Vvert}}}
\newcommand{\gA}{g_{_{\Avert}}}
\newcommand{\fVi}[1]{f^{#1}_{_{\Vvert}}}
\newcommand{\seai}[1]{a_{#1}}
\newcommand{\seapi}[1]{a'_{#1}}
\newcommand{\seAi}[2]{A_{#1}^{#2}}
\newcommand{\sewi}[1]{w_{#1}}
\newcommand{\seWi}[1]{W_{#1}}
\newcommand{\seWsi}[1]{W^{*}_{#1}}
\newcommand{\seWti}[2]{W_{#1}^{#2}}
\newcommand{\sewti}[2]{w_{#1}^{#2}}
\newcommand{\seSig}[1]{\Sigma_{#1}\lpar\sla{\pmom}\rpar}
\newcommand{\ww}{{\rm{w}}}
\newcommand{\ew}{{\rm{ew}}}
\newcommand{\ct}{{\rm{ct}}}
\newcommand{\nonSE}{\rm{non-SE}}
\newcommand{\leading}{\rm{\ssL}}
%
%
\newcommand{\bbff}[1]{{\overline B}_{#1}}
\newcommand{\sW}{p_{_W}}
\newcommand{\sZ}{p_{_Z}}
\newcommand{\ssp}{s_p}
\newcommand{\fW}{f_{_W}}
\newcommand{\fZ}{f_{_Z}}
\newcommand{\subMSB}[1]{{#1}_{\mbox{$\overline{\scriptscriptstyle MS}$}}}
\newcommand{\supMSB}[1]{{#1}^{\mbox{$\overline{\scriptscriptstyle MS}$}}}
\newcommand{\redMSB}{{\mbox{$\overline{\scriptscriptstyle MS}$}}}
\newcommand{\gpbb}{g'_{0}}
\newcommand{\Zconip}[1]{Z'_{#1}}
\newcommand{\bpff}[4]{B'_{#1}\lpar #2;#3,#4\rpar}             
\newcommand{\xidf}{\xi^2-1}
\newcommand{\tDdr}{1/{\bar{\varepsilon}}}
\newcommand{\cRz}{{\cal R}_{_Z}}
\newcommand{\cRg}{{\cal R}_{\gamma}}
\newcommand{\Sz}{\Sigma_{_Z}}
\newcommand{\alh}{{\hat\alpha}}
\newcommand{\alhz}{\alpha_{\ssZ}}
\newcommand{\Phzg}{{\hat\Pi}_{_{\zb\ab}}}
\newcommand{\fvvert}{F^{\rm vert}_{_V}}
\newcommand{\gavert}{G^{\rm vert}_{_A}}
\newcommand{\bmv}{{\overline m}_{_V}}
\newcommand{\Sgn}{\Sigma_{\gamma\hkn}}
\newcommand{\rmboxd}{{\rm Box}_d\lpar s,t,u;M_1,M_2,M_3,M_4\rpar}
\newcommand{\rmboxc}{{\rm Box}_c\lpar s,t,u;M_1,M_2,M_3,M_4\rpar}
%
%
\newcommand{\Afaci}[1]{A_{#1}}
\newcommand{\Afacis}[1]{A^2_{#1}}
\newcommand{\upar}[1]{u}
\newcommand{\upari}[1]{u_{#1}}
\newcommand{\vpari}[1]{v_{#1}}
\newcommand{\lpari}[1]{l_{#1}}
\newcommand{\Lpari}[1]{l_{#1}}
\newcommand{\Nff}[2]{N^{(#1)}_{#2}}
\newcommand{\Sff}[2]{S^{(#1)}_{#2}}
\newcommand{\sSff}{S}
\newcommand{\etafd}[2]{\eta_d\lpar#1,#2\rpar}
\newcommand{\sigdu}[2]{\sigma_{#1#2}}
\newcommand{\scalc}[4]{c_{0}\lpar #1;#2,#3,#4\rpar}
\newcommand{\scald}[2]{d_{0}\lpar #1,#2\rpar}
\newcommand{\scaldi}[3]{d_{0}^{#1}\lpar #2,#3\rpar}
\newcommand{\pir}[1]{\Pi^{\rm{\ssR}}\lpar #1\rpar}
\newcommand{\sigh}{\sigma_{\rm had}}
\newcommand{\dah}{\Delta\alpha^{(5)}_{\rm had}}
\newcommand{\dat}{\Delta\alpha_{\rm top}}
\newcommand{\Vqed}[3]{V_1^{\rm sub}\lpar#1;#2,#3\rpar}
\newcommand{\thetah}{{\hat\theta}}
\newcommand{\smlon}{\frac{\mlones}{s}}
\newcommand{\lntwo}{\ln 2}
\newcommand{\wmin}{w_{\rm min}}
\newcommand{\kmin}{k_{\rm min}}
\newcommand{\mdls}{\Big|}
\newcommand{\smf}{\frac{\mfs}{s}}
\newcommand{\bint}{\beta_{\rm int}}
\newcommand{\IRv}{V_{_{\rm IR}}}
\newcommand{\IRr}{R_{_{\rm IR}}}
\newcommand{\fssts}{\frac{s^2}{t^2}}
\newcommand{\fssus}{\frac{s^2}{u^2}}
\newcommand{\optM}{1+\frac{t}{M^2}}
\newcommand{\opuM}{1+\frac{u}{M^2}}
\newcommand{\ftM}{\lpar -\frac{t}{M^2}\rpar}
\newcommand{\fuM}{\lpar -\frac{u}{M^2}\rpar}
\newcommand{\omsM}{1-\frac{s}{M^2}}
\newcommand{\xsf}{\sigma_{_{\rm F}}}
\newcommand{\xsb}{\sigma_{_{\rm B}}}
\newcommand{\afb}{A_{_{\rm FB}}}
\newcommand{\rsoft}{\rm soft}
\newcommand{\rms}{\rm s}
\newcommand{\rsmx}{\sqrt{s_{\rm max}}}
\newcommand{\rspm}{\sqrt{s_{\pm}}}
\newcommand{\rsp}{\sqrt{s_{+}}}
\newcommand{\rsm}{\sqrt{s_{-}}}
\newcommand{\sigmx}{\sigma_{\rm max}}
\newcommand{\gG}[2]{G_{#1}^{#2}}
\newcommand{\gacomm}[2]{\lpar #1 - #2\gfd\rpar}
\newcommand{\fcsx}{\frac{1}{\ctwsix}}
\newcommand{\fcq}{\frac{1}{\ctwf}}
\newcommand{\fcs}{\frac{1}{\ctws}}
\newcommand{\affs}[2]{{\cal A}_{#1}\lpar #2\rpar}                   
\newcommand{\stwei}{s_{\theta}^8}
\def\mdan{\vspace{1mm}\mpar{\hfil$\downarrow$new\hfil}\vspace{-1mm}
          \ignorespaces}
\def\muan{\vspace{-1mm}\mpar{\hfil$\uparrow$new\hfil}\vspace{1mm}\ignorespaces}
\def\mlan{\vspace{-1mm}\mpar{\hfil$\rightarrow$new\hfil}\vspace{1mm}\ignorespaces}
\def\mnnew{\mpar{\hfil NEWNEW \hfil}\ignorespaces}
%
%
\newcommand{\boxc}[2]{{\cal{B}}_{#1}^{#2}}
\newcommand{\boxct}[3]{{\cal{B}}_{#1}^{#2}\lpar{#3}\rpar}
\newcommand{\hboxc}[3]{\hat{{\cal{B}}}_{#1}^{#2}\lpar{#3}\rpar}
\newcommand{\vev}{\langle v \rangle}
\newcommand{\vevi}[1]{\langle v_{#1}\rangle}
\newcommand{\vevs}{\langle v^2   \rangle}
\newcommand{\fwfrV}[5]{\Sigma_{_V}\lpar #1,#2,#3;#4,#5 \rpar}
\newcommand{\fwfrS}[7]{\Sigma_{_S}\lpar #1,#2,#3;#4,#5;#6,#7 \rpar}
\newcommand{\fSi}[1]{f^{#1}_{_{\Svert}}}
\newcommand{\fPi}[1]{f^{#1}_{_{\Pvert}}}
\newcommand{\mXs}{m_{_X}}
\newcommand{\mXss}{m^2_{_X}}
\newcommand{\mYs}{M^2_{_Y}}
\newcommand{\xik}{\xi_k}
\newcommand{\xiks}{\xi^2_k}
\newcommand{\mpls}{m^2_+}
\newcommand{\mmis}{m^2_-}
%
\newcommand{\SN}{\Sigma_{_N}}
\newcommand{\SC}{\Sigma_{_C}}
\newcommand{\SPN}{\Sigma'_{_N}}
\newcommand{\PFf}{\Pi^{\fer}_{_F}}
\newcommand{\PFb}{\Pi^{\bos}_{_F}}
\newcommand{\dPZ}{\Delta{\hat\Pi}_{_Z}}
\newcommand{\Sfin}{\Sigma_{_F}}
\newcommand{\Sfir}{\Sigma_{_R}}
\newcommand{\Sfinh}{{\hat\Sigma}_{_F}}
\newcommand{\Sfinf}{\Sigma^{\fer}_{_F}}
\newcommand{\Sfinbh}{\Sigma^{\bos}_{_F}}
\newcommand{\alf}{\alpha^{\fer}}
\newcommand{\alhfz}{\alpha^{\fer}\lpar{\mzs}\rpar}
\newcommand{\alhfs}{\alpha^{\fer}\lpar{\sman}\rpar}
\newcommand{\gfQ}{g^f_{_{Q}}}
\newcommand{\gfL}{g^f_{_{L}}}
\newcommand{\ccf}{\frac{\gbs}{16\,\pi^2}}
\newcommand{\chq}{{\hat c}^4}
\newcommand{\muuq}{m_{u'}}
\newcommand{\muus}{m^2_{u'}}
\newcommand{\mdd}{m_{d'}}
\newcommand{\clf}[2]{\mathrm{Cli}_{_#1}\lpar\displaystyle{#2}\rpar}
\def\stes{\sin^2\theta}
\def\acal{\cal A}
\def\alr{A_{_{\rm{LR}}}}
\newcommand{\barQ}{\overline Q}
\newcommand{\Sptg}{\Sigma'_{_{3Q}}}
\newcommand{\Sptt}{\Sigma'_{_{33}}}
\newcommand{\Ppgg}{\Pi'_{\ph\ph}}
\newcommand{\Pww}{\Pi_{_{\wb\wb}}}
\newcommand{\capV}[2]{{\cal F}^{#2}_{_{#1}}}
\newcommand{\bt}{\beta_t}
\newcommand{\btp}{\beta'_t}
\newcommand{\mhsix}{M^6_{_H}}
\newcommand{\topt}{{\cal T}_{33}}
\newcommand{\topq}{{\cal T}_4}
\newcommand{\Phzgf}{{\hat\Pi}^{\fer}_{_{\zb\ab}}}
\newcommand{\Phzgb}{{\hat\Pi}^{\bos}_{_{\zb\ab}}}
\newcommand{\Sfirh}{{\hat\Sigma}_{_R}}
\newcommand{\Szgh}{{\hat\Sigma}_{_{\zb\ab}}}
\newcommand{\Szghb}{{\hat\Sigma}^{\bos}_{_{\zb\ab}}}
\newcommand{\Szghf}{{\hat\Sigma}^{\fer}_{_{\zb\ab}}}
\newcommand{\Szgb}{\Sigma^{\bos}_{_{\zb\ab}}}
\newcommand{\Szgf}{\Sigma^{\fer}_{_{\zb\ab}}}
\newcommand{\chig}{\chi_{\ph}}
\newcommand{\chiz}{\chi_{\ssZ}}
\newcommand{\Sfih}{{\hat\Sigma}}
\newcommand{\Szzh}{\hat{\Sigma}_{_{\zb\zb}}}
\newcommand{\dPZf}{\Delta{\hat\Pi}^f_{_{\zb}}}
\newcommand{\khZdf}[1]{{\hat\kappa}^{#1}_{_{\zb}}}
\newcommand{\chf}{{\hat c}^4}
\newcommand{\amp}[2]{{\cal{A}}_{_{#1}}^{\rm{#2}}}
\newcommand{\hatvm}[1]{{\hat v}^-_{#1}}
\newcommand{\hatvp}[1]{{\hat v}^+_{#1}}
\newcommand{\hatvpm}[1]{{\hat v}^{\pm}_{#1}}
\newcommand{\kvz}[1]{\kappa^{\zb #1}_{_V}}
\newcommand{\barp}{\overline p}
\newcommand{\delw}{\Delta_{_{\wb}}}
\newcommand{\bdelw}{{\bar{\Delta}}_{_{\wb}}}
\newcommand{\bdelf}{{\bar{\Delta}}_{\ff}}
\newcommand{\delz}{\Delta_{_\zb}}
\newcommand{\deli}[1]{\Delta\lpar{#1}\rpar}
\newcommand{\hdeli}[1]{{\hat{\Delta}}\lpar{#1}\rpar}
\newcommand{\chizb}{\chi_{_\zb}}
\newcommand{\Swwp}{\Sigma'_{_{\wb\wb}}}
\newcommand{\epph}{\varepsilon'/2}
\newcommand{\sbffp}[1]{B'_{#1}}
\newcommand{\epss}{\varepsilon^*}
\newcommand{\Ddrhs}{{\ds\frac{1}{\hat{\varepsilon}^2}}}
\newcommand{\lnmsb}{L_{_\wb}}
\newcommand{\lnsmsb}{L^2_{_\wb}}
\newcommand{\tpni}{\lpar 2\pi\rpar^n\ib}
\newcommand{\tpn}{2^n\,\pi^{n-2}}
\newcommand{\cmf}{M_f}
\newcommand{\cmfs}{M^2_f}
\newcommand{\toDdr}{{\ds\frac{2}{{\bar{\varepsilon}}}}}
\newcommand{\troDdr}{{\ds\frac{3}{{\bar{\varepsilon}}}}}
\newcommand{\totDdr}{{\ds\frac{3}{{2\,\bar{\varepsilon}}}}}
\newcommand{\foDdr}{{\ds\frac{4}{{\bar{\varepsilon}}}}}
\newcommand{\smh}{m_{_H}}
\newcommand{\smhs}{m^2_{_H}}
\newcommand{\Ph}{\Pi_{_\hb}}
\newcommand{\Sphh}{\Sigma'_{_{\hb\hb}}}
\newcommand{\bh}{\beta}
\newcommand{\alsn}{\alpha^{(n_f)}_{_S}}
\newcommand{\smq}{m_q}
\newcommand{\smqp}{m_{q'}}
\newcommand{\shb}{h}
\newcommand{\hab}{A}
\newcommand{\hbpm}{H^{\pm}}
\newcommand{\hbp}{H^{+}}
\newcommand{\hbm}{H^{-}}
\newcommand{\msh}{M_h}
\newcommand{\mha}{M_{_A}}
\newcommand{\mhc}{M_{_{H^{\pm}}}}
\newcommand{\mshs}{M^2_h}
\newcommand{\mhas}{M^2_{_A}}
\newcommand{\barfp}{\overline{f'}}
\newcommand{\chiii}{{\hat c}^3}
\newcommand{\chiv}{{\hat c}^4}
\newcommand{\chv}{{\hat c}^5}
\newcommand{\chvi}{{\hat c}^6}
\newcommand{\alsvi}{\alpha^{6}_{_S}}
\newcommand{\tww}{t_{_W}}
\newcommand{\ti}{t_{1}}
\newcommand{\tii}{t_{2}}
\newcommand{\tiii}{t_{3}}
\newcommand{\tiv}{t_{4}}
\newcommand{\psla}{\hbox{\rlap/p}}
\newcommand{\qsla}{\hbox{\rlap/q}}
\newcommand{\nsla}{\hbox{\rlap/n}}
\newcommand{\lsla}{\hbox{\rlap/l}}
\newcommand{\msla}{\hbox{\rlap/m}}
\newcommand{\cnsla}{\hbox{\rlap/N}}
\newcommand{\clsla}{\hbox{\rlap/L}}
\newcommand{\cmsla}{\hbox{\rlap/M}}
\newcommand{\blmt}{\lrbr - 3\rrbr}
\newcommand{\blfo}{\lrbr 4 1\rrbr}
\newcommand{\bltp}{\lrbr 2 +\rrbr}
\newcommand{\clitwo}[1]{{\rm{Li}}_{2}\lpar{#1}\rpar}
\newcommand{\clitri}[1]{{\rm{Li}}_{3}\lpar{#1}\rpar}
\newcommand{\xt}{x_{\ft}}
\newcommand{\zt}{z_{\ft}}
\newcommand{\Ht}{h_{\ft}}
\newcommand{\xts}{x^2_{\ft}}
\newcommand{\zts}{z^2_{\ft}}
\newcommand{\Hts}{h^2_{\ft}}
\newcommand{\ztc}{z^3_{\ft}}
\newcommand{\Htc}{h^3_{\ft}}
\newcommand{\ztq}{z^4_{\ft}}
\newcommand{\Htq}{h^4_{\ft}}
\newcommand{\ztv}{z^5_{\ft}}
\newcommand{\Htv}{h^5_{\ft}}
\newcommand{\ztx}{z^6_{\ft}}
\newcommand{\Htx}{h^6_{\ft}}
\newcommand{\ztz}{z^7_{\ft}}
\newcommand{\Htz}{h^7_{\ft}}
\newcommand{\sht}{\sqrt{\Ht}}
\newcommand{\atan}[1]{{\rm{arctan}}\lpar{#1}\rpar}
\newcommand{\dbff}[3]{{\hat{B}}_{_{{#2}{#3}}}\lpar{#1}\rpar}
\newcommand{\ztbs}{{\bar{z}}^{2}_{\ft}}
\newcommand{\ztb}{{\bar{z}}_{\ft}}
\newcommand{\Htbs}{{\bar{h}}^{2}_{\ft}}
\newcommand{\Htb}{{\bar{h}}_{\ft}}
\newcommand{\Hztb}{{\bar{hz}}_{\ft}}
\newcommand{\Ln}[1]{{\rm{Ln}}\lpar{#1}\rpar}
\newcommand{\Lns}[1]{{\rm{Ln}}^2\lpar{#1}\rpar}
\newcommand{\wt}{w_{\ft}}
\newcommand{\wts}{w^2_{\ft}}
\newcommand{\wtb}{\overline{w}}
\newcommand{\fra}{\frac{1}{2}}
\newcommand{\frb}{\frac{1}{4}}
\newcommand{\frc}{\frac{3}{2}}
\newcommand{\frd}{\frac{3}{4}}
\newcommand{\fre}{\frac{9}{2}}
\newcommand{\frf}{\frac{9}{4}}
\newcommand{\frg}{\frac{5}{4}}
\newcommand{\frh}{\frac{5}{2}}
\newcommand{\fri}{\frac{1}{8}}
\newcommand{\frj}{\frac{7}{4}}
\newcommand{\frl}{\frac{7}{8}}
\newcommand{\Spzzh}{\hat{\Sigma}'_{_{\zb\zb}}}
\newcommand{\sqs}{\sqrt{s}}
\newcommand{\Rtg}{R_{_{3Q}}}
\newcommand{\Rtt}{R_{_{33}}}
\newcommand{\Rww}{R_{_{\wb\wb}}}
\newcommand{\ssA}{{\scriptscriptstyle{A}}}
\newcommand{\ssB}{{\scriptscriptstyle{B}}}
\newcommand{\ssC}{{\scriptscriptstyle{C}}}
\newcommand{\ssD}{{\scriptscriptstyle{D}}}
\newcommand{\ssE}{{\scriptscriptstyle{E}}}
\newcommand{\ssF}{{\scriptscriptstyle{F}}}
\newcommand{\ssG}{{\scriptscriptstyle{G}}}
\newcommand{\ssH}{{\scriptscriptstyle{H}}}
\newcommand{\ssI}{{\scriptscriptstyle{I}}}
\newcommand{\ssJ}{{\scriptscriptstyle{J}}}
\newcommand{\ssK}{{\scriptscriptstyle{K}}}
\newcommand{\ssL}{{\scriptscriptstyle{L}}}
\newcommand{\ssM}{{\scriptscriptstyle{M}}}
\newcommand{\ssN}{{\scriptscriptstyle{N}}}
\newcommand{\ssO}{{\scriptscriptstyle{O}}}
\newcommand{\ssP}{{\scriptscriptstyle{P}}}
\newcommand{\ssQ}{{\scriptscriptstyle{Q}}}
\newcommand{\ssR}{{\scriptscriptstyle{R}}}
\newcommand{\ssS}{{\scriptscriptstyle{S}}}
\newcommand{\ssT}{{\scriptscriptstyle{T}}}
\newcommand{\ssU}{{\scriptscriptstyle{U}}}
\newcommand{\ssV}{{\scriptscriptstyle{V}}}
\newcommand{\ssW}{{\scriptscriptstyle{W}}}
\newcommand{\ssX}{{\scriptscriptstyle{X}}}
\newcommand{\ssY}{{\scriptscriptstyle{Y}}}
\newcommand{\ssZ}{{\scriptscriptstyle{Z}}}
\newcommand{\ssWF}{\rm{\scriptscriptstyle{WF}}}
\newcommand{\OLA}{\rm{\scriptscriptstyle{OLA}}}
\newcommand{\QED}{\rm{\scriptscriptstyle{QED}}}
\newcommand{\QCD}{\rm{\scriptscriptstyle{QCD}}}
\newcommand{\EW}{\rm{\scriptscriptstyle{EW}}}
\newcommand{\UV}{\rm{\scriptscriptstyle{UV}}}
\newcommand{\IR}{\rm{\scriptscriptstyle{IR}}}
\newcommand{\OMS}{\rm{\scriptscriptstyle{OMS}}}
\newcommand{\GMS}{\rm{\scriptscriptstyle{GMS}}}
\newcommand{\sszero}{\rm{\scriptscriptstyle{0}}}
\newcommand{\DiagramFermionToBosonFullWithMomenta}[8][70]{
  \vcenter{\hbox{
  \SetScale{0.8}
  \begin{picture}(#1,50)(15,15)
    \put(27,22){$\nearrow$}
    \put(27,54){$\searrow$}
    \put(59,29){$\to$}
    \ArrowLine(25,25)(50,50)      \Text(34,20)[lc]{#6} \Text(11,20)[lc]{#3}
    \ArrowLine(50,50)(25,75)      \Text(34,60)[lc]{#7} \Text(11,60)[lc]{#4}
    \Photon(50,50)(90,50){2}{8}   \Text(80,40)[lc]{#2} \Text(55,33)[ct]{#8}
    \Vertex(50,50){2,5}          \Text(60,48)[cb]{#5}
    \Vertex(90,50){2}
  \end{picture}}}
  }
\newcommand{\DiagramFermionToBosonPropagator}[4][85]{
  \vcenter{\hbox{
  \SetScale{0.8}
  \begin{picture}(#1,50)(15,15)
    \ArrowLine(25,25)(50,50)
    \ArrowLine(50,50)(25,75)
    \Photon(50,50)(105,50){2}{8}   \Text(90,40)[lc]{#2}
    \Vertex(50,50){0.5}         \Text(80,48)[cb]{#3}
    \GCirc(82,50){8}{1}            \Text(55,48)[cb]{#4}
    \Vertex(105,50){2}
  \end{picture}}}
  }
\newcommand{\DiagramFermionToBosonEffective}[3][70]{
  \vcenter{\hbox{
  \SetScale{0.8}
  \begin{picture}(#1,50)(15,15)
    \ArrowLine(25,25)(50,50)
    \ArrowLine(50,50)(25,75)
    \Photon(50,50)(90,50){2}{8}   \Text(80,40)[lc]{#2}
    \BBoxc(50,50)(5,5)            \Text(55,48)[cb]{#3}
    \Vertex(90,50){2}
  \end{picture}}}
  }
\newcommand{\DiagramFermionToBosonFull}[3][70]{
  \vcenter{\hbox{
  \SetScale{0.8}
  \begin{picture}(#1,50)(15,15)
    \ArrowLine(25,25)(50,50)
    \ArrowLine(50,50)(25,75)
    \Photon(50,50)(90,50){2}{8}   \Text(80,40)[lc]{#2}
    \Vertex(50,50){2.5}          \Text(60,48)[cb]{#3}
    \Vertex(90,50){2}
  \end{picture}}}
  }
\newcommand{\expgw}{\frac{\gf\mws}{2\srt\,\pi^2}}
\newcommand{\expgz}{\frac{\gf\mzs}{2\srt\,\pi^2}}
\newcommand{\Spww}{\Sigma'_{_{\wb\wb}}}
\newcommand{\shf}{{\hat s}^4}
\newcommand{\acz}{\scff{0}}
\newcommand{\acoo}{\scff{11}}
\newcommand{\acod}{\scff{12}}
\newcommand{\acdo}{\scff{21}}
\newcommand{\acdd}{\scff{22}}
\newcommand{\acdt}{\scff{23}}
\newcommand{\acdq}{\scff{24}}
\newcommand{\acto}{\scff{31}}
\newcommand{\actd}{\scff{32}}
\newcommand{\actt}{\scff{33}}
\newcommand{\actq}{\scff{34}}
\newcommand{\actc}{\scff{35}}
\newcommand{\acts}{\scff{36}}
\newcommand{\acoA}{\scff{1A}}
\newcommand{\acdA}{\scff{2A}}
\newcommand{\acdB}{\scff{2B}}
\newcommand{\acdC}{\scff{2C}}
\newcommand{\acdD}{\scff{2D}}
\newcommand{\actA}{\scff{3A}}
\newcommand{\actB}{\scff{3B}}
\newcommand{\actC}{\scff{3C}}
\newcommand{\ada}{\sdff{0}}
\newcommand{\adb}{\sdff{11}}
\newcommand{\adc}{\sdff{12}}
\newcommand{\add}{\sdff{13}}
\newcommand{\ade}{\sdff{21}}
\newcommand{\adf}{\sdff{22}}
\newcommand{\adg}{\sdff{23}}
\newcommand{\adh}{\sdff{24}}
\newcommand{\adi}{\sdff{25}}
\newcommand{\adj}{\sdff{26}}
\newcommand{\adl}{\sdff{27}}
\newcommand{\adm}{\sdff{31}}
\newcommand{\adn}{\sdff{32}}
\newcommand{\ado}{\sdff{33}}
\newcommand{\adp}{\sdff{34}}
\newcommand{\adq}{\sdff{35}}
\newcommand{\adr}{\sdff{36}}
\newcommand{\ads}{\sdff{37}}
\newcommand{\adt}{\sdff{38}}
\newcommand{\adu}{\sdff{39}}
\newcommand{\adw}{\sdff{310}}
\newcommand{\adv}{\sdff{311}}
\newcommand{\ady}{\sdff{312}}
\newcommand{\adz}{\sdff{313}}
\newcommand{\admt}{\frac{\tman}{\sman}}
\newcommand{\admu}{\frac{\uman}{\sman}}
\newcommand{\bee}{\beta_{e}}
\newcommand{\beW}{\beta_{_\wb}}
\newcommand{\beWs}{\beta^2_{_\wb}}
\newcommand{\etaW}{\eta_{_\wb}}
\newcommand{\toDdrh}{{\ds\frac{2}{{\hat{\varepsilon}}}}}
\newcommand{\mhcub}{M^3_{_H}}
\newcommand{\afba}[1]{A^{#1}_{_{\rm FB}}}
\newcommand{\alra}[1]{A^{#1}_{_{\rm LR}}}
\newcommand{\etavz}[1]{\eta^{\zb #1}_{_V}}
\newcommand{\eilc}{\gamma}
\newcommand{\dalpha}{\Delta\alpha}
\newcommand{\drho}{\Delta\rho}
\newcommand{\drhov}{\delta\rho}
\newcommand{\dkapv}{\delta\kappa}
\newcommand{\drhovh}{\delta{\hat{\rho}}}
\newcommand{\drhovb}{\delta{\hat{\rho}}}
\newcommand{\sss}[1]{\scriptscriptstyle{#1}}
\newcommand{\epsb}{\bar\varepsilon}
\newcommand{\gll}{\Gamma_{\fl}}
\newcommand{\gqq}{\Gamma_{\fq}}
\newcommand{\Imsi}[1]{I^2_{#1}}
\newcommand{\reni}[1]{R_{#1}}
\newcommand{\renis}[1]{R^2_{#1}}
\newcommand{\sreni}[1]{\sqrt{R_{#1}}}
\newcommand{\dalphav}{\Delta\alpha^{(5)}(\mzs)}
\newcommand{\Rvaz}[1]{g^{#1}_{\sss{\zb}}}
\newcommand{\rvab}[1]{{\bar{g}}_{#1}}
\newcommand{\rvabs}[1]{{\bar{g}}^2_{#1}}
\newcommand{\rab }[1]{{\bar{a}}_{#1}}
\newcommand{\rvb }[1]{{\bar{v}}_{#1}}
\newcommand{\rabs}[1]{{\bar{a}}^2_{#1}}
\newcommand{\rva }[1]{g_{#1}}
\newcommand{\rvas}[1]{g^2_{#1}}
\newcommand{\Rva }[1]{G_{#1}}
\newcommand{\Rvac}[1]{G^{*}_{#1}}
\newcommand{\Rvah }[1]{{\hat{G}}_{#1}}
\newcommand{\Rvahc}[1]{{\hat{G}}^{*}_{#1}}
\newcommand{\Rvas}[1]{G^2_{#1}}
\newcommand{\rhobi} [1]{{\bar{\rho}}_{#1}}
\newcommand{\kappai}[1]{\kappa_{#1}}
\newcommand{\rhois}[1]{\rho^2_{#1}}
\newcommand{\rhohi} [1]{{\hat{\rho}}_{#1}}
\newcommand{\rhobpi}[1]{{\bar{\rho}}'_{#1}}
\newcommand{\dd}{{\mathrm d}}
\newcommand{\matrm}[1]{\mbox{\scriptsize #1}}
\newcommand{\vecc}[1]{\mbox{\boldmath $#1$}}
\newcommand{\bm}[1]{\mbox{\boldmath $#1$}}
\begin{center}
\parbox{15cm}{
\begin{center}
{\Large \bf Project ``CalcPHEP: \\ Calculus for Precision High Energy Physics''}
\end{center}
}
\end{center}

\begin{center}
{\large {\underline {D.Bardin}}, G.Passarino$^*$, L.Kalinovskaya, P.Christova, A.Andonov, 
S.Bondarenko$^{**}$ and G.Nanava}
\end{center}

\begin{center}
\parbox{10cm} {
\begin{center}
 {\it Laboratory of Nuclear Problems, JINR} \\
 {\it Joliot-Curie, 6, Dubna, Moscow region, Russia} \\
 fax: -7(096)21-66666, e-mail: {\tt bardin$@$nusun.jinr.ru}\\
 $^* {\mbox{\it Dipartimento di Fisica Teorica, Universit\`a di Torino, Italy}}$\\
 {\it INFN, Sezione di Torino, Italy }\\
 $^{**}$ {\it Bogoluobov Laboratory of Theoretical Physics, JINR}
\end{center} }
\end{center}
\footnoterule
\noindent
{\footnotesize \noindent
Work supported by INTAS N$^{o}$00-00313
and by the European Union under contract HPRN-CT-2000-00149.}
\section{Introduction}
The {\tt CalcPHEP} collaboration joins the efforts of several groups of theorists
known very well in the field of theoretical support of various experiments in HEP, 
particularly at SLAC and LEP, 
(see, for instance \cite{Bardin:1995aa}, \cite{Bardin:PCP} and \cite{Kobel:LEP2000}).
The first phase of the {\tt CalcPHEP} system was realized in the site 
{\it http://brg.jinr.ru/} in 2000--2001.
It is written mostly in {\tt FORM3}, \cite{Vermaseren:2000f}. 
In this talk, we will describe the present status
and our plans for the realization of next phases of the {\tt CalcPHEP} project
aimed at the theoretical support of experiments at modern and future accelerators:
TEVATRON, LHC, electron Linear Colliders (LC's) i.e. TESLA, NLC, CLIC, and muon factories.
Within this project, we are creating a four-level computer system which eventually 
must automatically calculate pseudo- and realistic observables for more and more 
complicated processes 
of elementary particle interactions, using the principle of knowledge storing. 
Upon completion of the second phase of the project, started January 2002
with duration of about three years, we plan to have a complete set of computer codes, 
accessible via 
an Internet-based environment and realizing the complete chain of calculations 
``from the Lagrangian to the realistic distributions'' at the one-loop level 
precision including all $1\to2$ decays, $2\to2$ processes and certain classes
of $2\to3$ processes.

\subsection{CalcPHEP group}
The {\tt CalcPHEP} group was formed in 2001 in sector N$^{o}1$ NEOVP LJAP. 

During the first phase of the project in 2000--2001, the {\tt CalcPHEP} group 
created the site {\tt brg.jinr.ru}, where the development in two strategic 
directions is foreseen:

\vspace{-3mm}

\begin{enumerate}
\item Creation of a softwear product, capable to compute HEP observables with 
one-loop precision for complicated processes 
of elementary particle interactions, using the principle of knowledge storing.
Application: LHC.

\vspace{-3mm}

\item Works towards two-loop precision level control of simple processes:
$1\to 2$, $1\to 3$ and $2\to 2$. Application: GigaZ option of electron LC's. 
\end{enumerate}

\subsection{A little bit of history}
There are two historical {\bf sources} of {\tt CalcPHEP} project:\\
{\bf 1.} From one side it roots back to many codes written by Dubna group 
aimed at a theoretical support of HEP experiments in the past:\\
{\bf 1975 -- 1986:} support of CERN DIS experiments (BCDMS, EMC, NMC), creation of program 
{\tt TERAD}; support of CERN neutrino experiments (CHARM-I, CDHSW and CHARM-II), creation of
programs {\tt NUDIS, INVMUD, NUFITTER}.\\
{\bf 1983 -- 1989:} Foundation of the DZRCG --- 
``Dubna--Zeuthen Radiative Correction Group'', creation of EW library {\tt DIZET}; 
creation of the program {\tt ZBIZON} --- the fore-runner of {\tt ZFITTER}~\cite{Bardin:1999yd}.\\
{\bf 1989 -- 1997:} support of the DIS experiments at HERA, creation of the program 
{\tt HECTOR}; participation in SMC experiment at CERN with the program ${\mu}{\bf{e}}{\it{la}}$.\\
{\bf 1989 -- 2001:} Theoretical support of experiments at LEP, SLC
(DELPHI, L3, ALEPH, OPAL and SLD).\\
{\bf 2.} From the other side, a monograph 
``The Standard Model in the Making'' was written~\cite{Bardin:1999ak}.
While working on the book, the authors wrote hundreds of ``book-supporting'' 
form-codes, which comprised the proto-type of future {\tt CalcPHEP} system.

Like well known codes of LEP era: {\tt TOPAZ0}~\cite{Montagna:1999kp}, 
{\tt ZFITTER}~\cite{Bardin:1999yd}, ${\cal{KK}}$MC~\cite{KKMC},
{\tt CalcPHEP} is supposed to be a tool for precision calculations
of pseudo- and realistic observables. Let's remind these definitions that arose
in depth of LEP community:

\newtheorem{guess}{Definition}
\begin{guess}
Realistic Observables are the (differential) cross-sections (more general event
distributions) for a reaction, e.g.
\bqa
\fep\fem\rightarrow (\ph,\zb) \rightarrow\ff\barf(n\ph)
\nn
\eqa
calculated with all available in the literature higher order
corrections (QCD, EW), including real and virtual QED 
photonic corrections, possibly accounting for kinematical cuts.
\end{guess}
\begin{guess}
Pseudo-Observables are related to measured quantities
by some de-convo\-lu\-tion or unfolding procedure (e.g. undressing
of QED corrections).
The concept itself of pseudo-observability is rather difficult to define.
One way say that the experiments measure some
primordial distributions which are then reduced to secondary quantities under
some set of specific assumptions (definitions).
\end{guess}
$\zb$ decay partial width represents typical example of pseudo-observables,
i.e. it has to be {\em defined}.
At the tree level, we define it
as a quantity described by the square of one diagram:
\vspace*{-7mm}
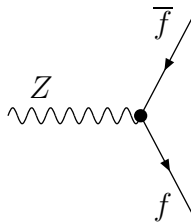
\[
\ba{c}
\begin{picture}(125,86)(0,15)
  \Photon(0,43)(50,43){3}{7}
    \Vertex(50,43){2.5}
  \ArrowLine(70,81)(50,43)
  \ArrowLine(50,43)(70,5)
\Text(55,72)[lb]{$\fbf$}
\Text(12.5,50)[bc]{$\zb$}
\Text(55,14)[lt]{$\ff$}
\end{picture}
\ea
\]
\section{LEP, Precision High Energy Physics and its Future}
One may say that during recent years a new physical discipline was born.
We call it PHEP, Precision High Energy Physics. Experimentally, it finally
shaped in the result of glorious 12 year LEP era: measurements at $\zb$ resonance
in {\bf 1989 -- 1995}, and reaching an unprecedented experimental accuracy $\leq 10^{-3}$, 
and measurements above $\zb$ resonance in {\bf 1995 -- 2000}, at higher energies, 
where high enough experimental accuracy was also reached $\leq1\%$. 
By 2/11/2000 LEP2 possibly saw hints of ``God blessed'' particle --- Higgs boson, but was stopped, 
unfortunately, mainly due to lack of financing.

For the first time huge HEP facility challenged for theoreticians to perform 
calculations with uncertainty better than experimental errors of
$\ord{10^{-3}}$ and, eventually, efforts of many groups of theoreticians 
allowed the achievement of the theoretical precision 
of the order $2.5\cdot 10^{-4}$ at the $Z$ resonance and $2-3\cdot 10^{-4}$ 
at LEP2 energies.

This, in turn, greatly contributed to the success of
precision tests of the SM, the main result of LEP era,
which laid the foundation of the Precision 
High Energy Physics. This is why our project got this suffix {\tt PHEP}.  

\subsection{Future of PHEP}
PHEP has good perspectives and after the end of LEP.
Several Input parameters of the Standard Model (SM) are expected to be improved in near future.

Recent discrepancy in the muon amm:
\bqa
a^{\sss{SM}}_{\mu}&=&116 591 661(114)\times 10^{-11}
\nl
a^{\sss{EXP}}_{\mu}(Average)&=&116 592 023(151)
\nl
a^{\sss{EXP}}_{\mu}-a^{\sss{SM}}_{\mu}&=&362(189)\quad 2\sigma\mbox{~difference}
\nl
&&\mbox{exp. error (151) should be improved soon up to $\sim(50)$,}
\eqa
necessitates an improvement of the knowledge of the hadronic contribution of
$\dalpha^{(5)}_{\had}\lpar\mzs\rpar$ to the running e.m. coupling.
An experimental input for
$\sigma\lpar\fep\fem\to\mbox{hadrons}\rpar$ at cms energies (1-4 GeV)
is expected from BES-II, BEPC (Beijing), VEPP2000 (Novossibirsk) and
DAFNE at cms energies around $\phi$-meson.

Very important should be projected improvements of mass measurements: $\mw$, $\mt$.

LEP1 finished with {\em indirect} result for the top mass: $\mt=169^{+10}_{-8}$ GeV;
while LEP1 $\oplus$ TEVATRON constraint yielded: $\mt=174.5^{+4.4}_{-4.2}$ GeV.

LEP2 reached for $\wb$ mass $\mw$: $\mw=80.450\pm 0.039$, in the {\em direct} 
measurements and $\mw=80.373\pm 0.023$ as {\em indirect} result. 

TEVATRON in RUN-I reached: $\mw=80.454\pm 0.060$ GeV, $\mt=174.3\pm 5.1$ GeV.

Much better precision tags are expected to be reached at 
TEVATRON, RUN-II (recently started): 
$\Delta\mw\sim 20$MeV, $\Delta\mt\sim 2$ GeV;
and later at LHC (not so sooner than in 2006, however):
$\Delta\mw\sim 15$ MeV, $\Delta\mt\sim 1$ GeV.

Where, when and with which mass Higgs boson might be discovered?\\
$-$ TEVATRON has a serious chance to see Higgs up to mass 180 GeV;
however it will require very high integrated luminosity: $\int{\cal{L}} \geq 5fb^{-1}$;
\\
$-$ LHC, will cover all allowed mass range up to 500 GeV 
(not so soon, after 2007);\\ 
$-$ LC's and muon factories (after 2010--2012).

New horizon of PHEP will be opened with experiments at electron LC's: 
TESLA (DESY) particularly with GigaZ option, i.e. coming back to $Z$ resonance with 
statistics $10^9$; 
CLIC (CERN); JLC (KEK), NLC(SLAC, LNBL, LLNL, FNAL) and Muon Storage Rings 
(Higgs Factory) --- all that more than in ten years from now. 

One expects fantastic precision tags there in:\\
$-$ $\Delta\sin^2\theta_{eff}\sim 0.00002$;\\
$-$ $\Delta\mw\sim 6$ MeV, $\Delta\mt\sim 100-200$MeV;\\
$-$ $\Delta\mh\sim 100$ MeV (from $\fep\fem\to\zb\hb$);\\
$-$ and detail study of Higgs boson properties.

Given our LEP1 experience one should definitely state that
{\bf\em 2-loop precision level control will be absolutely necessary
for the analysis of these data!}

{\it One may conclude that PHEP has a bright future:
all future colliders --- TEVATRON, LHC, electron 
LC's (TESLA, NLC, CLIC) and muon factories will be, actually, PHEP facilities!
For data analysis, they will surely require qualitatively new 
level of both theoretical predictions and principally new computer codes.}

\section{Necessary notion}
In order to understand the language of {\tt CalcPHEP} one has to introduce many
notions and notations.
\subsection{Input Parameter Set, IPS}
The Minimal Standard Model (MSM), contains large number of Input Parameters:\\
$25=$ 2 interaction constants $\alpha$ and $\als$ \\
$\phantom{.}\quad\oplus$ 8 mixing angles (CKM and possible lepton analogs) \\
$\phantom{.}\quad\oplus$ 15 masses (12 fundamental fermions and 3 fundamental bosons
$\zb,\,\wb,\,\hb$).

However, the number 25 is {\em minimal}.
MSM is {\bf unable} to compute its IPS from first principles;
MSM is {\bf able} to compute {\em any observable}
$O^{\rm exp}_{i}$ in terms of its IPS:
\bqa
O^{\rm exp}_{i}\lpar\mbox{measured}\rpar&\leftrightarrow&
O^{\rm theor}_{i}\lpar\mbox{calculated, as a function of IPS}\rpar.
\eqa
This is the way how precision measurements set {\em constraints} on IPS.
\subsubsection{Number of free parameters in fits of $\zb$ resonance observables}
At $\zb$ resonance, not all 25 parameters matter. Actually only 5 parameters:
\bqa
\dalpha^{(5)}_{\had}\lpar\mzs\rpar,\qquad\als\lpar\mzs\rpar,\qquad{\mt}\,,
\qquad\mz\,,\qquad\mh\,,
\eqa
which we call the Standard LEP1 IPS, matter. 

Using $\mz$, measured at $\zb$ peak itself with the precision
$\sim2\times10^{-5}$, and also reach information from the other measurements for:
\bqa
\als\lpar\mzs\rpar,\qquad\mt\,,\qquad\mw\,,
\eqa
we {\em approach} one-parameter fit, with
Higgs boson mass $\mh$ being the only fitted parameter.
The result of such a fit was shown in the {\em Blue band} figure, the most
celebrated LEP era figure, derived with the aid of {\tt TOPAZ0}~\cite{Montagna:1999kp} 
and {\tt ZFITTER}~\cite{Bardin:1999ak} codes. 

\subsection{Quantum Fields of the SM} 
Here we sketch all fundamental quantum fields of the SM in one of the most
general gauges --- $\Rxi$, with three arbitrary gauge parameters
$\gparA,\,\gparZ,\,\gpar$.\\
\leftline{\underline{Three generation of fermions or matter fields:}}
\[
\Fermionline{0}
\qquad
\ff\quad=\quad\lcbr\quad
\ba{c}
\lpar
\ba{c}
\fnu \\ \fl
\ea
\rpar
\quad=\\ \\ 
\lpar
\ba{c}
\fU \\ \fD
\ea
\rpar
\quad=
\ea
\quad
\ba{c}
\lpar
\ba{c}
\fnue \\ \fem
\ea
\rpar
\\ \\ 
\lpar
\ba{c}
\fu \\ \fd
\ea
\rpar
\ea
\quad
\ba{c}
\lpar
\ba{c}
\fnum \\ \flm
\ea
\rpar
\\ \\ 
\lpar
\ba{c}
\fc \\ \fs
\ea
\rpar
\ea
\quad
\ba{c}
\lpar
\ba{c}
\fnut \\ \flt
\ea
\rpar
\\ \\ 
\lpar
\ba{c}
\ft \\ \ffb
\ea
\rpar
\ea
\right.
\]

\noindent
possess masses, $\mf$, charges, $\qf$,
and third projections of weak isospin, $\tcif$:
\[
\mf,
\quad
\qf=
\ba{c}
\lpar
\ba{cccc}
\fnu&\quad\fl&\quad\fU&\quad\fD\\ \\
  0 & -1 &{\ds{+\frac{2}{3}}}&{\ds{-\frac{1}{3}}}
\ea
\rpar,
\ea
\quad
\tcif=
\ba{c}
\lpar
\ba{cccc}
\fnu&\quad\fl&\quad\fU&\quad\fD\\ \\
{\ds{+\frac{1}{2}}} & {\ds{-\frac{1}{2}}} &
{\ds{+\frac{1}{2}}} & {\ds{-\frac{1}{2}}}
\ea
\rpar.
\ea
\]

\leftline{\underline{Gauge fields:}}
\[
\ba{lll}
\mbox{Vector~bosons}&\mbox{Unphysical~scalars}\quad
&\mbox{Faddeev--Popov ghosts} \\[2mm]
\Photonline{0}\quad\ab \quad &
                       \quad &\Ghostline{0}\quad\fpyA     \\ \\
\Zbosline{0}  \quad\zb\;(\mz)\quad&
\Philine{0}   \quad\hkn \quad&\Ghostline{0}\quad\fpyZ     \\ \\
\Wbosline{0}  \quad\wbpm(\mw)\quad&
\Phicline{0}  \quad\hkpm\quad&\Ghostline{0}\quad\fpxpm    \\ \\
\mbox{Gluon} & & \\ 
\mbox{possesses strong interaction} & &                   \\[2mm]
\Gluonline{0} \quad\glu\quad&&\Ghostline{0}\quad\fpyG     \\ 
\ea
\]
{possess physical charges and physical masses}

\hspace{5.5cm}{possess physical charges and unphysical masses}

\hspace{10cm}{and unphysical charges.}

\leftline{\underline{Higgs field:}}

\vspace*{3mm}

\lefteqn{\Philine{0}\quad\hb\;(\mh)\quad\mbox{is a scalar,~neutral,~massive field.}}
\subsubsection{The Lagrangian in $\Rxi$ gauge, Feynman Rules}
At the ground level of {\tt CalcPHEP} system one has this Lagrangian
\bqa
{\cal{L}}={\cal{L}}(\mbox{IPS of 25 parameters, 17 fields, 3 gauge parameters}),
\eqa
from which one derives {\em primary Feynman rules for vertices}.

\subsubsection{Propagators in $\Rxi$ gauge}
Here we list propagators in $\Rxi$ gauge, the other important bricks of {\tt CalcPHEP} 
system.\\
\underline{Propagator of a fermion, $\ff$}:
\[
\ba{cc}
\ba{c}
\Fermionline{6.}
\\ \ff
\ea
&\qquad
\ds{\frac{-\ib\sla{\pmom}+\mf}{\pmoms+\mfs}}\;
\ea
\]
\underline{Vector boson propagators:}
\[
\ba{ccl}
\ab&\;
\Photonline{0}
&\qquad
\ds{
\frac{1}{\pmoms}
\lcbr
\drii{\mu}{\nu}+\lpar\gparAs-1\rpar\frac{\pmomi{\mu}\pmomi{\nu}}{\pmoms}
\rcbr }
\\[1mm]
\zb&\;
\Zbosline{0}
&\qquad
\ds{
\propf{\pmom}{\mzs}
\lcbr
\drii{\mu}{\nu}+\lpar\gparZs-1\rpar
\frac{\pmomi{\mu}\pmomi{\nu}}{\pmoms+\gparZs\mzs}
\rcbr \qquad }
\\[1mm]
{\wbpm}
&\;
\Wbosline{0}
&\qquad
\ds{
\propf{\pmom}{\mws}
\lcbr
\drii{\mu}{\nu}+\lpar\gpars-1\rpar 
\frac{\pmomi{\mu}\pmomi{\nu}}{\pmoms+\gpars\mws}
\rcbr }
\ea
\]
\vskip 5pt
\noindent
\underline{Propagators of unphysical fields:}
\vspace*{-10mm}
\[
\ba{cccc}
&
&\qquad
\ba{c}
\Ghostline{6.}
\\[1mm] \fpyA
\ea
&\quad
\ds{\frac{\gparA}{\pmoms}}\;
\\[1mm]
\ba{c}
\Philine{6.}
\\[1mm] \hkn
\ea
&\quad
\ds{\frac{1}{\pmoms+\gparZs\mzs}}\;,
&\qquad
\ba{c}
\Ghostline{6.}
\\[1mm] \fpyZ
\ea
&\quad
\ds{\frac{\gparZ}{\pmoms+\gparZs\mzs}}\;
\\[1mm]
\ba{c}
\Phicline{6.}
\\[1mm] \hkpm
\ea
&\quad
\ds{\frac{1}{\pmoms+\gpars\mws}}\;,
&\qquad\quad
\ba{c}
\Ghostline{6.}
\\[1mm] \fpxpm
\ea
&\quad
\ds{\frac{\gpar}{\pmoms+\gpars\mws}}\;
\ea\quad
\]
\vskip 5pt
\noindent
\underline{Propagator of the physical scalar field, $\hb$-boson}
\[
\ba{cc}
\ba{c}
\Philine{6.}
\\ \hb
\ea
&\qquad
\ds{\frac{1}{\pmoms+\mhs}}\;
\ea
\]
\subsection{Scalar $\saff{0}$, $\sbff{0}$, etc functions} 
For calculation of one-loop integrals {\tt CalcPHEP} uses the standard scalar $\saff{0}$, 
$\sbff{0}$, $\scff{0}$ and $\sdff{0}$ functions~\cite{Bardin:1999ak}.

\leftline{\em One-point integrals or $\saff{0}$ functions, are met in tadpoles diagrams:}
\[
  \vcenter{\hbox{
  \begin{picture}(120,80)(0,0)
  \Line(0,50)(50,50)
  \CArc(75,50)(25,-180,180)
  \Text(110,50)[cb]{$\mlone$}
  \end{picture}}}
\vspace*{-10mm}
\]
We give its defining expression:
\bqa
\ib\pi^2\,\aff{0}{\mlone} = \tHs^{\fmon} \intmomi{n}{\imom}\, 
\propa{\imom}{\mlone}\;,
\eqa
and the answer in the {\em dimensional regularization}:
\bqa
\aff{0}{\mlone}=\mlones\lpar - \Ddr - 1 + \lmass{\frac{\mlones}{\tHss}}\rpar
+ \ord{\dre}\;,
\eqa
where the ultraviolet pole is:
\bqa
\Ddr = \frac{2}{\dre} - \eilc - \lpi\,,\qquad n=4-\dre\,.
\eqa
\leftline{\it Two-point integrals or $\sbff{0}$-functions are met in self-energy 
diagrams:}
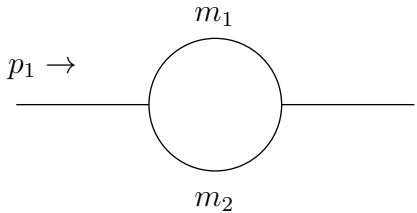
\[
  \vcenter{\hbox{
  \begin{picture}(150,90)(0,0)
  \Line(0,50)(50,50)
  \CArc(75,50)(25,-180,180)
  \Line(100,50)(150,50)
  \Text(10,60)[cb]{$\pone\to$}
  \Text(75,80)[cb]{$\mone$}
  \Text(75,10)[cb]{$\mtwo$}
  \end{picture}}}
\vspace*{-5mm}
\]
We limit ourselves by giving its defining expression:
\bqa
&&\ib\pi^2\bff{0}{\pones}{\mone}{\mtwo}= \tHs^{\fmon} \intmomi{n}{\imom}
                                       \frac{1}{\dpropi{0}\dpropi{1}}\;,
\nl
&&\dpropi{0}= \imoms+\mones-\ib\ep,
\qquad
\dpropi{1} =\lpar\imom+\pone\rpar^2+\mtwos-\ib\ep\,.
\eqa
\leftline{\it Three-point integrals, $\scff{}$ functions, are met in vertices:}
\vspace*{-2mm}
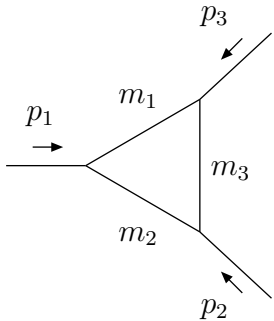
\[
  \vcenter{\hbox{
  \begin{picture}(140,130)(-15,-15)
    \Line(30,50)(73,75)        \Text(50,72)[cb]{$\mone$}
    \Line(73,75)(73,25)        \Text(50,20)[cb]{$\mtwo$}
    \Line(73,25)(30,50)        \Text(85,45)[cb]{$\mtre$}
    \Line(0,50)(30,50)    
    \Line(100,100)(73,75)   
    \Line(100,0)(73,25)   
    \LongArrow(10,57)(20,57)   \Text(13,65)[cb]{$\pone$}
    \LongArrow(89,98)(82,91)   \Text(79,103)[cb]{$\ptre$}
    \LongArrow(89,2)(82,9)     \Text(79,-8)[cb]{$\ptwo$}
  \end{picture}}}
\]
Its defining expression reads:
\bq
\ib\pi^2\,\cff{0}{\pones}{\ptwos}{\gmv}{\mone}{\mtwo}{\mtre} = \tHs^{\fmon} 
\intmomi{n}{\imom}\frac{1}{\dpropi{0}\dpropi{1}\dpropi{2}}\;,
\eq
\bq
\dpropi{0}=\imoms+\mones-\ib\ep,
\quad
\dpropi{1}=\lpar\imom+\pone \rpar^2+\mtwos-\ib\ep,
\quad
\dpropi{2}=\lpar\imom+\pone+\ptwo \rpar^2+\mtres-\ib\ep,
\eq
where
$\gmv=\lpar\pone+\ptwo\rpar^2$ is one of the Mandelstamm variables:
$\sman,\tman$ or $\uman$. 

\leftline{\it Four-point integrals, $\sdff{}$-functions, are met in boxes.}

Presently, {\tt CalcPHEP} knows ALL about reduction of up to four-point functions 
up to third rank tensors and of the so-called {\it special} functions, 
which are due to peculiar form of the photonic propagator in the $\Rxi$ gauge, see
\cite{Bardin:1999ak}.

\subsection{Processes in the SM}
One should be aware of a hierarchical classification of processes accepted in 
{\tt CalcPHEP} and of a relevant notion of {\em independent structures}, 
or {\em independent amplitude form factors}, which
number is deeply related to the number of {\em independent helicity amplitudes}
by which a process may be described (below we present these numbers for unpolarized
cases). 

\subsubsection{Decays~ $1\to 2$}

There are $B\to\ff\barf$ and $3B$ decays:
\begin{itemize}
\item[$\bullet$] $\hb\to\ff\barf\quad\,$ \hspace{2.2cm} (one structure)
\item[$\bullet$] $\zb\to\ff\barf$, $(\ph\to\ff\barf)\qquad\;$ (three structures)
\item[$\bullet$] $\wb\to\ff\bar{\ffp}$, $(\ft\to\wbp\ffb)\quad$ (four structures)
\item[$\bullet$] $\hb\to\zb\zb,\;\wbp\wbm\quad$
\item[$\bullet$] $\zb\to\wbp\wbm$
\end{itemize}

\subsubsection{Processes~ $2\to 2$}
 
There are $2\ff\to 2\ff$ processes, which in turn are subdivided into Neutral Current (NC)
and Charged Current (CC) ones: 
\begin{itemize}
\item[$\bullet$] NC:~ $\ff\fbf\to(\ph,\zb,\hb)\to\ffp\overline{\ffp}\quad$ (4,6)~10 structures
depending on whether initial and final state fermion masses are ignored
\item[$\bullet$] CC:~ $\ffi{1}\bffi{2}\to(\wb)\to\ffi{3}\bffi{4}\quad$
\end{itemize}

Next, there are many processes of a kind $\vb\ff\to \ffp\vb'$, in particular
\begin{itemize}
\item[$\bullet$] compton-effect: $\ph\fe\to\ph\fe$,~~$\zb\to\ff\barf\ph$
\item[$\bullet$] $\fep\fem\to\wbp\wbm,\;\zb\zb,\;\zb\ph,\ph\ph$
\end{itemize}

Decays~$1\to 3$ are cross-channels of the previous processes and their one-loop 
description in terms of independent objects, mentioned above, one gets for free.
Present level of {\tt CalcPHEP} has a lot of preparations for all above processes,
but far not all is put into the working areas of the site {\tt brg.jinr.ru}.

\subsubsection{Processes~ $2\to 3$}
They comprise a very reach family, for instance:
\begin{itemize}
\item[$\bullet$] $\fep\fem\to(\ph,\zb,\hb)\to\ff\barf\ph$~.
\end{itemize}
Their implementation is one of main goals of the second phase of {\tt CalcPHEP} project.
Corresponding decays~ $1\to 4$ are again  cross-channels of the previous processes and
need not be studied separately.

\subsubsection{Processes~ $2\to 4$}
To this family belongs $4$ fermion processes of LEP2.
Their study is not foreseen at the second phase of {\tt CalcPHEP} project,
but might be a subject of its third phase.
\section{Building Blocks and knowledge storing}
\subsection{Simplest decay: $\zb\to\ff\fbf$}
\subsubsection{Amplitude of $\zb\to\ff\fbf$ decay at tree level}
Its tree level diagram was already presented at the end of Section 1.2;
the corresponding amplitude reads:
\bq
\Vveri{\mu}{\zb\ff\barf} =
\tpfi\,{\ds\frac{\ib\gb}{2\,\cow}}\,\gadu{\mu}\,
\biggl[\tcif\bigl( 1+\gfd\bigr) - 2\qf\siws\biggr],
\label{Ztree}
\eq
with vector and axial coupling constants:
$\vc{\ff}=\tcif-2\qf\siws\,,\;\ac{\ff}=\tcif.$ Note appearance of the two structures
in \eqn{Ztree}, which might be termed as $L$ and $Q$ structures, correspondingly.
Note also, that \eqn{Ztree} as well as all below, are written in Pauli metrics that is
used by {\tt CalcPHEP}.

\subsubsection{Amplitude of $\zb\to\ff\fbf$ decay with loop corrections}
It might be schematically depicted as a sum of one-loop vertices and counter terms:
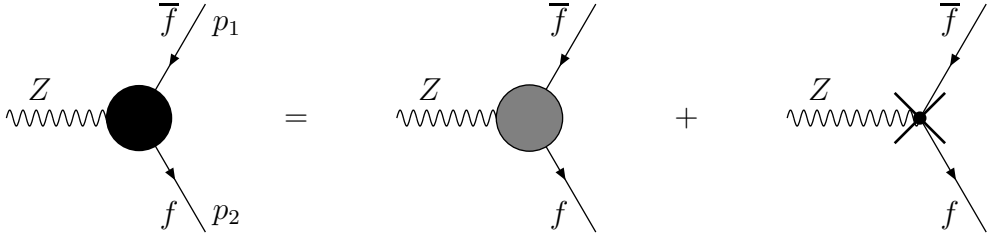
\begin{figure}[ht]
\vspace*{-20mm}
\[
\ba{ccccc}
\begin{picture}(95,86)(0,40)
  \Photon(0,43)(50,43){3}{10}
    \Vertex(50,43){12.5}
  \ArrowLine(75,86)(50,43)
  \ArrowLine(50,43)(75,0)
\Text(58,74)[lb]{$\fbf$}
\Text(78,74)[lb]{$\pone$}
\Text(12.5,50)[bc]{$\zb$}
\Text(58,12)[lt]{$\ff$}
\Text(78,10)[lt]{$\ptwo$}
\end{picture}
&=&\qquad
\begin{picture}(95,86)(0,40)
  \Photon(0,43)(50,43){3}{10}
  \ArrowLine(75,86)(50,43)
  \ArrowLine(50,43)(75,0)
  \GCirc(50,43){12.5}{0.5}
\Text(58,74)[lb]{$\fbf$}
\Text(12.5,50)[bc]{$\zb$}
\Text(58,12)[lt]{$\ff$}
\end{picture}
&+&\qquad
\begin{picture}(95,86)(0,40)
  \Photon(0,43)(50,43){3}{10}
\SetScale{2.0}
    \Line(20.25,16.75)(29.75,26.25)
    \Line(20.25,26.25)(29.75,16.75)
\SetScale{1.0}
  \ArrowLine(75,86)(50,43)
    \Vertex(50,43){2.5}
  \ArrowLine(50,43)(75,0)
\Text(58,74)[lb]{$\fbf$}
\Text(12.5,50)[bc]{$\zb$}
\Text(58,12)[lt]{$\ff$}
\end{picture}
\ea
\]
\vspace{5mm}
\end{figure}

\noindent
In the most general case (but for unpolarized study)
the one-loop amplitude may be parameterized by the three {\em scalar form factors}:
\bq
\Vveri{\mu}{\zb\ff\barf} =
\tpfi\,\frac{\gbc}{16\,\pi^2\,2\,\cow}\,\gadu{\mu}\,
\biggl[
\ib\tcif F_{\ssL}\gdp - 2\ib\qf\siws F_{\ssQ}+\mf\lpar\pone-\ptwo\rpar F_{\sss{D}}
\biggr].
\label{Zloop}
\eq
Given similarity of \eqnsc{Ztree}{Zloop}, the latter is called sometimes 
Improved Born Approximation (IBA) amplitude. 
\subsubsection{QED diagrams and corrections}
The QED diagrams comprise gauge invariant subsets, this is why they are considered 
sometimes separately:
\vspace*{-15mm}

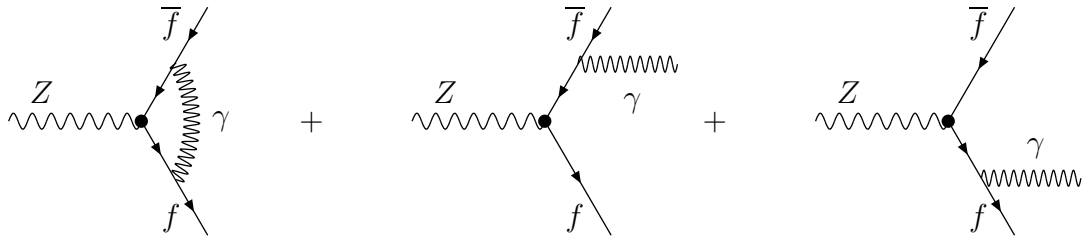
\[
\ba{ccccc}
\begin{picture}(100,86)(0,40)
  \Photon(0,43)(50,43){3}{7}
    \Vertex(50,43){2.5}
  \PhotonArc(25,43)(44,-30,30){3}{15}
  \ArrowLine(75,86)(62.5,64.5)
  \ArrowLine(62.5,64.5)(50,43)
  \ArrowLine(50,43)(62.5,21.5)
  \ArrowLine(62.5,21.5)(75,0)
\Text(58,74)[lb]{$\fbf$}
\Text(12.5,50)[bc]{$\zb$}
\Text(58,12)[lt]{$\ff$}
\Text(77,43)[lc]{$\ph$}
\end{picture}
&+&\qquad
\begin{picture}(100,86)(0,40)
  \Photon(0,43)(50,43){3}{7}
    \Vertex(50,43){2.5}
  \ArrowLine(75,86)(62.5,64.5)
  \ArrowLine(62.5,64.5)(50,43)
  \ArrowLine(50,43)(75,0)
  \Photon(62.5,64.5)(100,64.5){3}{10}
\Text(58,74)[lb]{$\fbf$}
\Text(12.5,50)[bc]{$\zb$}
\Text(58,12)[lt]{$\ff$}
\Text(80,50)[lc]{$\ph$}
\end{picture}
&+&\qquad
\begin{picture}(100,86)(0,40)
  \Photon(0,43)(50,43){3}{7}
    \Vertex(50,43){2.5}
  \ArrowLine(50,43)(62.5,21.5)
  \ArrowLine(62.5,21.5)(75,0)
  \ArrowLine(75,86)(50,43)
  \Photon(62.5,21.5)(100,21.5){3}{10}
\Text(58,74)[lb]{$\fbf$}
\Text(12.5,50)[bc]{$\zb$}
\Text(58,12)[lt]{$\ff$}
\Text(80,33)[lc]{$\ph$}
\end{picture}
\ea
\]
\vspace*{10mm}

\noindent
Their contribution to the partial $Z$ widths, in the case when no photon cuts are imposed,
reads: 
\bqa
\gff^{\rm{QED}}=\gff\lpar 1+\frac{3\alpha}{4\pi}\qfs\rpar.
\eqa
\subsection{Process $\fep\fem\to\ff\fbf$}
Coming to a more complicated case of a $2\ff\to2\ff$ process, we will illustrate
how building blocks, derived for a study of a lower level process, might be use at 
a higher level. 
\subsubsection{Tree-level diagrams and amplitudes of $\fep\fem\to\ff\fbf$}
Consider first the two tree-level diagrams with $\ph$ and $\zb$ exchanges in
order to introduce basis of relevant structures.

\vspace{-15mm}
\[
\ba{ccc}
\begin{picture}(125,86)(0,40)
  \Photon(25,43)(100,43){3}{15}
  \ArrowLine(125,86)(100,43)
    \Vertex(100,43){2.5}
  \ArrowLine(100,43)(125,0)
  \ArrowLine(0,0)(25,43)
    \Vertex(25,43){2.5}
  \ArrowLine(25,43)(0,86)
\Text(14,74)[lb]{$\fep$}
\Text(102,74)[lb]{$\fbf$}
\Text(62.5,50)[bc]{$\ab$}
\Text(14,12)[lt]{$\fem$}
\Text(102,12)[lt]{$\ff$}
\end{picture}
\qquad
&+&
\qquad
\begin{picture}(125,86)(0,40)
  \Photon(25,43)(100,43){3}{7}
  \ArrowLine(125,86)(100,43)
    \Vertex(100,43){2.5}
  \ArrowLine(100,43)(125,0)
  \ArrowLine(0,0)(25,43)
    \Vertex(25,43){2.5}
  \ArrowLine(25,43)(0,86)
\Text(14,74)[lb]{$\fep$}
\Text(102,74)[lb]{$\fbf$}
\Text(62.5,50)[bc]{$\zb$}
\Text(14,12)[lt]{$\fem$}
\Text(102,12)[lt]{$\ff$}
\end{picture}
\ea
\]
\vspace*{5mm}

\bqa
\amp{\ph}{Born}&=&\frac{\ecs\qe\qf}{\sman}\gadu{\mu}\otimes\gadu{\mu}\,,
\nl[1mm]
\amp{\zb}{Born}&=&\frac{\ecs}{4\siws\cows}
\chi_{\ssZ}(\sman)\,
\gadu{\mu}\bigl(\vc{\fe}+\ac{\fe}\gfd\bigr)
\otimes
\gadu{\mu}\bigl(\vc{\ff}+\ac{\ff}\gfd\bigr)
\nl[1mm]
               &=&\frac{\ecs}{4\siws\cows}
\chi_{\ssZ}(\sman)\,
\gadu{\mu}\lrbr \tcie\gdp-2\qe\siws\rrbr
\otimes
\gadu{\mu}\lrbr \tcif\gdp-2\qf\siws\rrbr,
\eqa
where $\;\gdpm=1\pm\gfd\;$ and symbol $\otimes$ stands for a short-hand notation
\bqa
\gadu{\mu}\lpar\vc{1}+\ac{1}\gfd\rpar\otimes
\gadu{\nu}\lpar\vc{2}+\ac{2}\gfd\rpar&=&
\iap{\pmomi{+}}\gadu{\mu}\lpar\vc{1}+\ac{1}\gfd\rpar\ip{\pmomi{-}}
\op{\qmomi{-}}\gadu{\nu}\lpar\vc{2}+\ac{2}\gfd\rpar\oap{\qmomi{+}}
\nn
\eqa
and
\vspace*{-10mm}

\bqa
\chi_{\ssZ}(\sman) = \frac{1}{\sman-\mzs+\ib\sman\gz/\mz}\;.
\eqa
This amplitude is characterized by four structures:
\bq
LL=\gadu{\mu}\gdp\otimes\gadu{\mu}\gdp\,,\quad
LQ=\gadu{\mu}\gdp\otimes\gadu{\mu}\,,\quad
QL=\gadu{\mu}\otimes\gadu{\mu}\gdp\,,\quad
QQ=\gadu{\mu}\otimes\gadu{\mu}\,.
\eq
\subsubsection{One-loop amplitude for $\fep\fem\to\ff\fbf$}
``Dressed'' with one-loop vertices and counterterms, the $\ph$ and $\zb$ exchanges may 
be symbolically depicted as:
\vspace*{-15mm}
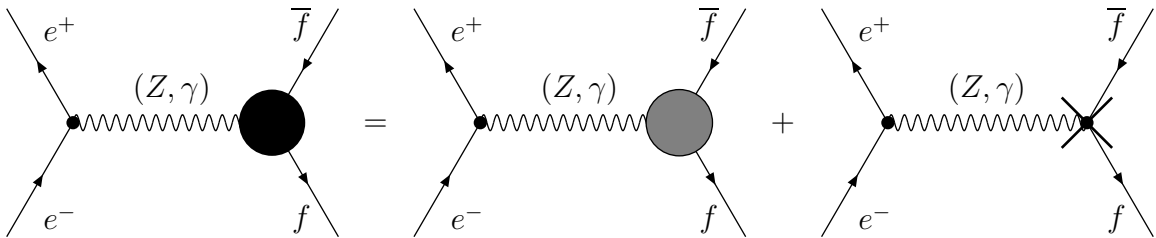
\[
\ba{ccccc}
\begin{picture}(125,86)(0,40)
  \Photon(25,43)(100,43){3}{15}
    \Vertex(100,43){12.5}
  \ArrowLine(125,86)(100,43)
  \ArrowLine(100,43)(125,0)
  \ArrowLine(0,0)(25,43)
    \Vertex(25,43){2.5}
  \ArrowLine(25,43)(0,86)
\Text(14,74)[lb]{$\fep$}
\Text(108,74)[lb]{$\fbf$}
\Text(62.5,50)[bc]{$(\zb,\ph)$}
\Text(14,12)[lt]{$\fem$}
\Text(108,12)[lt]{$\ff$}
\end{picture}
&=&
\begin{picture}(125,86)(0,40)
  \Photon(25,43)(100,43){3}{15}
  \ArrowLine(125,86)(100,43)
  \ArrowLine(100,43)(125,0)
    \GCirc(100,43){12.5}{0.5}
  \ArrowLine(0,0)(25,43)
  \ArrowLine(25,43)(0,86)
    \Vertex(25,43){2.5}
\Text(14,74)[lb]{$\fep$}
\Text(108,74)[lb]{$\fbf$}
\Text(62.5,50)[bc]{$(\zb,\ph)$}
\Text(14,12)[lt]{$\fem$}
\Text(108,12)[lt]{$\ff$}
\end{picture}
&+&
\begin{picture}(125,86)(0,40)
  \Photon(25,43)(100,43){3}{15}
\SetScale{2.0}
    \Line(45.25,16.75)(54.75,26.25)
    \Line(45.25,26.25)(54.75,16.75)
\SetScale{1.0}
  \ArrowLine(125,86)(100,43)
    \Vertex(100,43){2.5}
  \ArrowLine(100,43)(125,0)
  \ArrowLine(0,0)(25,43)
    \Vertex(25,43){2.5}
  \ArrowLine(25,43)(0,86)
\Text(14,74)[lb]{$\fep$}
\Text(108,74)[lb]{$\fbf$}
\Text(62.5,50)[bc]{$(\zb,\ph)$}
\Text(14,12)[lt]{$\fem$}
\Text(108,12)[lt]{$\ff$}
\end{picture}
\ea
\]

\clearpage

And similarly for the initial state vertex:
\vspace{-15mm}
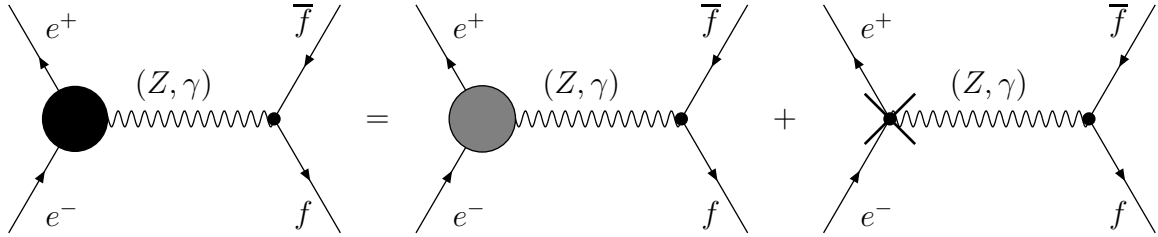
\[
\ba{ccccc}
\begin{picture}(125,86)(0,40)
  \Photon(25,43)(100,43){3}{15}
    \Vertex(25,43){12.5}
  \ArrowLine(0,0)(25,43)
  \ArrowLine(25,43)(0,86)
  \ArrowLine(125,86)(100,43)
    \Vertex(100,43){2.5}
  \ArrowLine(100,43)(125,0)
\Text(14,74)[lb]{$\fep$}
\Text(108,74)[lb]{$\fbf$}
\Text(62.5,50)[bc]{$(\zb,\ph)$}
\Text(14,12)[lt]{$\fem$}
\Text(108,12)[lt]{$\ff$}
\end{picture}
&=&
\begin{picture}(125,86)(0,40)
  \Photon(25,43)(100,43){3}{15}
  \ArrowLine(0,0)(25,43)
  \ArrowLine(25,43)(0,86)
    \GCirc(25,43){12.5}{0.5}
  \ArrowLine(125,86)(100,43)
    \Vertex(100,43){2.5}
  \ArrowLine(100,43)(125,0)
\Text(14,74)[lb]{$\fep$}
\Text(108,74)[lb]{$\fbf$}
\Text(62.5,50)[bc]{$(\zb,\ph)$}
\Text(14,12)[lt]{$\fem$}
\Text(108,12)[lt]{$\ff$}
\end{picture}
&+&
\begin{picture}(125,86)(0,40)
  \Photon(25,43)(100,43){3}{15}
\SetScale{2.0}
    \Line(7.75,16.75)(17.25,26.25)
    \Line(7.75,26.25)(17.25,16.75)
\SetScale{1.0}
    \Vertex(25,43){2.5}
  \ArrowLine(0,0)(25,43)
  \ArrowLine(25,43)(0,86)
  \ArrowLine(125,86)(100,43)
    \Vertex(100,43){2.5}
  \ArrowLine(100,43)(125,0)
\Text(14,74)[lb]{$\fep$}
\Text(108,74)[lb]{$\fbf$}
\Text(62.5,50)[bc]{$(\zb,\ph)$}
\Text(14,12)[lt]{$\fem$}
\Text(108,12)[lt]{$\ff$}
\end{picture}
\ea
\]
\vspace{10mm}

Where one can easily recognize building blocks already known from the calculation 
of one-loop radiative corrections for $\zb$ decay, however, now we need to dress
$\ph\to\ff\barf$ vertex too and add into consideration ``dressing'' of propagators: 
\vspace*{-15mm}
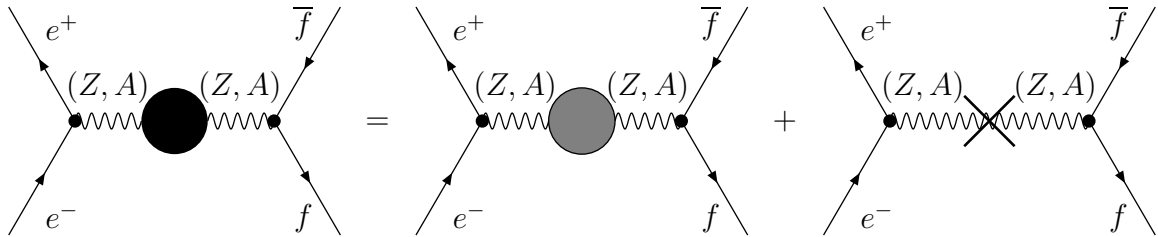
\[
\ba{ccccc}
\begin{picture}(125,86)(0,40)
  \Photon(25,43)(50,43){3}{5}
    \Vertex(62.5,43){12.5}
  \Photon(75,43)(100,43){3}{5}
  \ArrowLine(125,86)(100,43)
    \Vertex(100,43){2.5}
  \ArrowLine(100,43)(125,0)
  \ArrowLine(0,0)(25,43)
    \Vertex(25,43){2.5}
  \ArrowLine(25,43)(0,86)
\Text(14,74)[lb]{$\fep$}
\Text(108,74)[lb]{$\fbf$}
\Text(37.5,50)[bc]{$(\zb,\ab)$}
\Text(87.5,50)[bc]{$(\zb,\ab)$}
\Text(14,12)[lt]{$\fem$}
\Text(108,12)[lt]{$\ff$}
\end{picture}
&=&
\begin{picture}(125,86)(0,40)
  \Photon(25,43)(50,43){3}{5}
  \Photon(75,43)(100,43){3}{5}
  \GCirc(62.5,43){12.5}{0.5}
  \ArrowLine(125,86)(100,43)
    \Vertex(100,43){2.5}
  \ArrowLine(100,43)(125,0)
  \ArrowLine(0,0)(25,43)
    \Vertex(25,43){2.5}
  \ArrowLine(25,43)(0,86)
\Text(14,74)[lb]{$\fep$}
\Text(108,74)[lb]{$\fbf$}
\Text(37.5,50)[bc]{$(\zb,\ab)$}
\Text(87.5,50)[bc]{$(\zb,\ab)$}
\Text(14,12)[lt]{$\fem$}
\Text(108,12)[lt]{$\ff$}
\end{picture}
&+&
\begin{picture}(125,86)(0,40)
  \Photon(25,43)(100,43){3}{15}
\SetScale{2.0}
    \Line(26.5,16.75)(36,26.25)
    \Line(26.5,26.25)(36,16.75)
\SetScale{1.0}
  \ArrowLine(0,0)(25,43)
    \Vertex(25,43){2.5}
  \ArrowLine(25,43)(0,86)
  \ArrowLine(125,86)(100,43)
    \Vertex(100,43){2.5}
  \ArrowLine(100,43)(125,0)
\Text(14,74)[lb]{$\fep$}
\Text(108,74)[lb]{$\fbf$}
\Text(37.5,50)[bc]{$(\zb,\ab)$}
\Text(87.5,50)[bc]{$(\zb,\ab)$}
\Text(14,12)[lt]{$\fem$}
\Text(108,12)[lt]{$\ff$}
\end{picture}
\ea
\]
\vspace*{10mm}

To complete calculations of one-loop EWRC for the process $\fep\fem\to\ff\fbf$
one should add $\wb\wb$ and $\zb\zb$ boxes:
\vspace{-10mm}
\[
\ba{ccc}
\begin{picture}(100,70)(0,33)
  \ArrowLine(0,10)(25,10)
  \ArrowLine(25,10)(25,60)
  \ArrowLine(25,60)(0,60)
  \Photon(25,10)(75,10){3}{10}
  \Photon(25,60)(75,60){3}{10}
  \Vertex(25,10){2.5}
  \Vertex(25,60){2.5}
  \Vertex(75,10){2.5}
  \Vertex(75,60){2.5}
  \ArrowLine(100,60)(75,60)
  \ArrowLine(75,60)(75,10)
  \ArrowLine(75,10)(100,10)
  \Text(12.5,80)[tc]{$\fep$}
  \Text(50,80)[tc]{$\wb$}
  \Text(87.5,82)[tc]{$\fbd$}
  \Text(12.5,35)[lc]{$\fnue$}
  \Text(85,35)[rc]{$\fu$}
  \Text(12.5,-10)[cb]{$\fem$}
  \Text(50,-12)[bc]{$\wb$}
  \Text(87.5,-10)[cb]{$\fd$}
\end{picture}
\quad
&+&
\quad
\begin{picture}(100,70)(0,33)
  \ArrowLine(0,10)(25,10)
  \ArrowLine(25,10)(25,60)
  \ArrowLine(25,60)(0,60)
  \Photon(25,10)(75,60){3}{14}
  \Photon(25,60)(75,10){3}{14}
  \Vertex(25,10){2.5}
  \Vertex(25,60){2.5}
  \Vertex(75,10){2.5}
  \Vertex(75,60){2.5}
  \ArrowLine(100,60)(75,60)
  \ArrowLine(75,60)(75,10)
  \ArrowLine(75,10)(100,10)
  \Text(12.5,80)[tc]{$\fep$}
  \Text(50,70)[tc]{$\wb$}
  \Text(87.5,80)[tc]{$\fbu$}
  \Text(12.5,35)[lc]{$\fnue$}
  \Text(85,35)[rc]{$\fd$}
  \Text(12.5,-10)[cb]{$\fem$}
  \Text(50,0)[bc]{$\wb$}
  \Text(87.5,-10)[cb]{$\fu$}
\end{picture}
\ea
\]

\[
\ba{ccc}
\begin{picture}(100,70)(0,33)
  \ArrowLine(0,10)(25,10)
  \ArrowLine(25,10)(25,60)
  \ArrowLine(25,60)(0,60)
  \Photon(25,10)(75,10){3}{10}
  \Photon(25,60)(75,60){3}{10}
  \Vertex(25,10){2.5}
  \Vertex(25,60){2.5}
  \Vertex(75,10){2.5}
  \Vertex(75,60){2.5}
  \ArrowLine(100,60)(75,60)
  \ArrowLine(75,60)(75,10)
  \ArrowLine(75,10)(100,10)
  \Text(12.5,80)[tc]{$\fep$}
  \Text(50,80)[tc]{$(\zb,\ph)$}
  \Text(87.5,82)[tc]{$\fbf$}
  \Text(15,35)[lc]{$\fe$}
  \Text(85,35)[rc]{$\ff$}
  \Text(12.5,-10)[cb]{$\fem$}
  \Text(50,-12)[bc]{$(\zb,\ph)$}
  \Text(87.5,-12)[cb]{$\ff$}
\end{picture}
\quad
&+&
\quad
\begin{picture}(100,70)(0,33)
  \ArrowLine(0,10)(25,10)
  \ArrowLine(25,10)(25,60)
  \ArrowLine(25,60)(0,60)
  \Photon(25,10)(75,60){3}{14}
  \Photon(25,60)(75,10){3}{14}
  \Vertex(25,10){2.5}
  \Vertex(25,60){2.5}
  \Vertex(75,10){2.5}
  \Vertex(75,60){2.5}
  \ArrowLine(100,60)(75,60)
  \ArrowLine(75,60)(75,10)
  \ArrowLine(75,10)(100,10)
  \Text(12.5,80)[tc]{$\fep$}
  \Text(50,80)[tc]{$(\zb,\ph)$}
  \Text(87.5,82)[tc]{$\fbf$}
  \Text(15,35)[lc]{$\fe$}
  \Text(85,35)[rc]{$\ff$}
  \Text(12.5,-10)[cb]{$\fem$}
  \Text(50,-12)[bc]{$(\zb,\ph)$}
  \Text(87.5,-12)[cb]{$\ff$}
\end{picture}
\ea
\]
\vspace*{10mm}

$\zb\zb$, $\ph\ph$ and $\zb\ph$ boxes comprise  
gauge invariant subset of diagrams. Moreover, $\ph\ph$ and $\zb\ph$ boxes QED, vertices
and QED bremsstrahlung for NC $2\ff\to2\ff$ processes often are separated into  
a gauge invariant QED subset of diagrams.

Virtual QED one-loop diagrams together with four
QED bremsstrahlung diagrams form an Infra-Red Divergence (IRD) free subset.

This example clearly shows how the principle of knowledge storing is implemented 
within {\tt CalcPHEP} project: one starts from the simplest decays and collects all
relevant building blocks, BB's (off-shell with respect to boson mass). Then one moves to 
next level of complexity where all BB's computed at the previous level are requested,
but on top one needs more complicated objects (here boxes).

This strategy was realized in our recent calculations of the EWRC to the 
$e^{+}e^{-}\to f\bar{f}$ process, which are completely done with the aid of 
{\tt CalcPHEP}
system~\cite{BKN}. There is another study accomplished with {\tt CalcPHEP}~\cite{APV}.
\section{Status of the project}
Before discussing what is already available at the site {\tt brg.jinr.ru}, we present
some general information about {\tt CalcPHEP} system.
\subsection{Basic information about {\tt CalcPHEP}, {\bf keywords}}
{\tt CalcPHEP} is {\bf four-level computer system} for automatic calculation of 
pseudo- and realistic observables (decay rates, event distributions) for more and more 
complicated processes of elementary particle interactions, using the principle of 
knowledge storing. 

At each of the four levels there are:
\vspace*{-2mm}

\begin{enumerate}
\item Codes (written in FORM3), realizing full chain of analytic calculations  
from the SM Lagrangian ${\cal{L}}_{\sss{SM}}$ to the Ultra Violet Free Amplitudes, UVFA,
parameterized by a minimal set of scalar form factors; 
\vspace*{-3mm}

\item Codes (written in FORM3), realizing analytic calculations of a minimal subset of 
Helicity Amplitudes, HA's, followed by an automatic procedure of generation of codes for
numerical calculations of HA's (presently {\tt FORTRAN} codes, and in a near perspective
$C^{++}$ codes). 
\vspace*{-3mm}

\item Codes, realizing the so-called ``infrared rearrangement'' of HA's. This is needed
if the multiple photon emission is being exponentiated at the amplitude level. Currently,
bremsstrahlung photons are added in the lowest order and the third level is skipped.
\vspace*{-8mm}

\item Codes, that use HA's derived at the second (or third) level together with 
tree-level HA's for one-photon (or multiple-photon) emission, within a Monte Carlo event
generator, which is supposed to compute realistic distributions (presently {\tt FORTRAN}
codes, and in a near perspective $C^{++}$ codes.)
\end{enumerate}
\vspace*{-3mm}

It is an {\bf Internet based} and  {\bf Database based} system. The latter means 
that there is a storage of source codes written in different languages,
which talk to each other. They are placed into a homogeneous environment written in JAVA.

It follows {\bf Intermediate access principle}
i.e. full chain ``from the Lagrangian to realistic distribution'' should work out 
completely in real time, if someone requests this, 
however, it is supposed to have several ``entries'', say after each level,
or just providing the user with its final product --- a Monte Carlo event generator.

\subsection{Some technical data about {\tt CalcPHEP}}
\begin{enumerate}
\item Address {\it http://brg.jinr.ru/}
\vspace*{-3mm}

\item For realization of the site one used:
\vspace*{-3mm}

\begin{itemize}
\item[$-$] Apache web server under Linux,
\vspace*{-1mm}

\item[$-$] {\tt form3} compiler,
\vspace*{-1mm}

\item[$-$] mySQL server for relational databases.
\vspace*{-2mm}

\end{itemize}
\item In the current version, user-interface is realized with the use of PHP.
\vspace*{-3mm}
\item Nowadays, everything is being rewritten in JAVA in order to reach better 
``interactivity''
and to use reach possibilities of already written in this language libraries.\\
{\bf Main goal of this rewriting is to create a homogeneous environment both for 
accessing our codes from the database and for offering a possibility for simultaneous
work of several members of the group and external users.}
\end{enumerate}

\subsection{Present and nearest versions of {\tt CalcPHEP} system}
In 2001, we released two test-versions of {\tt CalcPHEP}:
\begin{enumerate}
\item {\bf v0.01 from March'01} realizes analytic calculations of one-loop UVFA 
for decays~$1\to 2$ (level-1).
\vspace*{-2.5mm}

\item {\bf v0.02d from March'02} returns numbers for one-loop decay widths
(levels-1,2) via temporary bypass of level 4. It realizes also levels-1,2 for 
$2\ff\to2\ff$ NC process.
\vspace*{-2.5mm}

\item One has very many almost finished ``preparations'' for the other processes
$2\to 2$ and decays $1\to 3$ (level-1). All this should comprise {\bf v0.03} of
Summer 2002.
\vspace*{-2.5mm}

\item An active work is being realized on implementation of level-4 for decays~$1\to 2$,
this should complete full chain ``from the SM Lagrangian to pseudo-observables'' for
the simplest decays.
\vspace*{-2.5mm}

\item There are many problems to be solved at the second or later phases of the project.
Among them one should mention:\\
-- automatic generation of Feynman Rules from a Lagrangian, \\
-- automatic generation of topologies of Feynman diagrams,  \\
-- graphical representation of the results.
\end{enumerate}

\section{Conclusion}
At a Symposium in honor of Professor Alberto Sirlin's 70th Birthday
was said: {\em A new frontier is as the horizon: most likely it is goodbye
to the one man show. Running a new Radiative Correction
project will be a little like running an experiment~\cite{SirlinGP}.}

Indeed, projects of such a kind as {\tt CalcPHEP} are definitely long term projects.
Remember, that {\tt ZFITTER} took about 12 years, about the same time exists  
already {\tt FeynArts}~\cite{Hahn:2000jm}.

Our nearest goal is the realization of the second phase of the project  
upon completion of which we plan to have a complete software product,
accessible via an Internet-based environment, and realizing the chain of calculations 
``from the Lagrangian to the realistic distributions'' at the 
one-loop level precision including some processes $2\to 3$ and decays $1\to 4$.
Plans also assume to perform an $R\&D$ for the third phase of the project
(see also~\cite{Tkachov:1996} and successive developments 
of~\cite{SirlinDB}--\cite{Passarino:2001wv}) which should begin in 2004.

Second phase is basically oriented on a common work of theoreticians of the Dubna group 
and the Knoxville--Krakow collaboration~\cite{KKMC}.

United group proposes to realize in 2002-2004 an important phase of CalcPHEP project:
oriented toward a merger of analytic results to be produced by Dubna team with MC 
event generators to be developed by Knoxville--Krakow collaboration\footnote{\normalsize
In this connection it is necessary to emphasize that any future 
code aimed at a comparison of experimental data with theory predictions should 
be a MC generator, since the processes at very high energies will have 
multi-particle final states that make impossible 
a semi-analytic approach used at LEP within {\tt ZFITTER} project.}.

Among most important milestones of first year, one should mention:
realization of the levels 2-4 for the simplest $Z(H,W)\to f\bar{f}$ decays;
completion of level 1 for the radiative $Z$ decay, $\zb\to f\bar{f}\gamma,$
work on which is already under way;
completion of levels 2-4 for the radiative $Z$ decay.

\end{document}